\documentclass[twoside,twocolumn,9pt]{article}
\usepackage{extsizes}
\usepackage[super,sort&compress,comma]{natbib} 
\usepackage[version=3]{mhchem}
\usepackage[left=1.5cm, right=1.5cm, top=1.785cm, bottom=2.0cm]{geometry}
\usepackage{balance}
\usepackage{times,mathptmx}
\usepackage{sectsty}
\usepackage{graphicx} 
\usepackage{lastpage}
\usepackage[format=plain,justification=justified,singlelinecheck=false,font={stretch=1.125,small,sf},labelfont=bf,labelsep=space]{caption}
\usepackage{float}
\usepackage{fancyhdr}
\usepackage{fnpos}
\usepackage[english]{babel}
\addto{\captionsenglish}{%
  
}
\usepackage{array}
\usepackage{droidsans}
\usepackage{charter}
\usepackage[T1]{fontenc}
\usepackage[usenames,dvipsnames]{xcolor}
\usepackage{setspace}
\usepackage[compact]{titlesec}
\usepackage{hyperref}

\usepackage{lipsum}
\usepackage{cuted}
\usepackage[version=3]{mhchem}
\usepackage[exponent-product = \cdot]{siunitx}

\usepackage{epstopdf}

\usepackage{cleveref}
\crefname{equation}{Eq.}{Eqs.}
\crefname{figure}{Fig.}{Figs.}
\crefrangeformat{equation}{Eqs.~#3(#1)#4--#5(#2)#6}

\usepackage{xfrac}
\usepackage{amssymb}

\definecolor{cream}{RGB}{222,217,201}

\begin{document}

\pagestyle{fancy}
\thispagestyle{plain}
\fancypagestyle{plain}{

\renewcommand{\headrulewidth}{0pt}
}

\makeFNbottom
\makeatletter
\renewcommand\LARGE{\@setfontsize\LARGE{15pt}{17}}
\renewcommand\Large{\@setfontsize\Large{12pt}{14}}
\renewcommand\large{\@setfontsize\large{10pt}{12}}
\renewcommand\footnotesize{\@setfontsize\footnotesize{7pt}{10}}
\renewcommand\scriptsize{\@setfontsize\scriptsize{7pt}{7}}
\makeatother

\renewcommand{\thefootnote}{\fnsymbol{footnote}}
\renewcommand\footnoterule{\vspace*{1pt}%
\color{cream}\hrule width 3.5in height 0.4pt \color{black} \vspace*{5pt}} 
\setcounter{secnumdepth}{5}

\makeatletter 
\renewcommand\@biblabel[1]{#1}            
\renewcommand\@makefntext[1]%
{\noindent\makebox[0pt][r]{\@thefnmark\,}#1}
\makeatother 
\renewcommand{\figurename}{\small{Fig.}~}
\sectionfont{\sffamily\Large}
\subsectionfont{\normalsize}
\subsubsectionfont{\bf}
\setstretch{1.125} 
\setlength{\skip\footins}{0.8cm}
\setlength{\footnotesep}{0.25cm}
\setlength{\jot}{10pt}
\titlespacing*{\section}{0pt}{4pt}{4pt}
\titlespacing*{\subsection}{0pt}{15pt}{1pt}

\fancyfoot{}
\fancyfoot[RO]{\footnotesize{\sffamily{1--\pageref{LastPage} ~\textbar  \hspace{2pt}\thepage}}}
\fancyfoot[LE]{\footnotesize{\sffamily{\thepage~\textbar\hspace{0cm} 1--\pageref{LastPage}}}}
\fancyhead{}
\renewcommand{\headrulewidth}{0pt} 
\renewcommand{\footrulewidth}{0pt}
\setlength{\arrayrulewidth}{1pt}
\setlength{\columnsep}{6.5mm}
\setlength\bibsep{1pt}

\makeatletter 
\newlength{\figrulesep} 
\setlength{\figrulesep}{0.5\textfloatsep} 

\newcommand{\topfigrule}{\vspace*{-1pt}%
\noindent{\color{cream}\rule[-\figrulesep]{\columnwidth}{1.5pt}} }

\newcommand{\botfigrule}{\vspace*{-2pt}%
\noindent{\color{cream}\rule[\figrulesep]{\columnwidth}{1.5pt}} }

\newcommand{\dblfigrule}{\vspace*{-1pt}%
\noindent{\color{cream}\rule[-\figrulesep]{\textwidth}{1.5pt}} }

\makeatother

\twocolumn[
  \begin{@twocolumnfalse}
\vspace{0cm}
\sffamily
\begin{tabular}{m{13.5cm} }

\noindent\LARGE{\textbf{Self-Thermoelectrophoresis at Low Salinity}} \\
\vspace{0.3cm} \\

\noindent\large{Joost de Graaf$^{\ast}$\textit{$^{a\ddag}$} and Sela Samin\textit{$^{a\ddag}$}} \\

\end{tabular}

 \end{@twocolumnfalse} \vspace{0.6cm}

  ]

\renewcommand*\rmdefault{bch}\normalfont\upshape
\rmfamily
\section*{}
\vspace{-1cm}


\footnotetext{\textit{$^{a}$~Institute for Theoretical Physics, Center for Extreme Matter and Emergent Phenomena, Utrecht University, Princetonplein 5, 3584 CC Utrecht, The Netherlands; E-mail: j.degraaf@uu.nl}}

\footnotetext{$\ddag$ Author contributions: Numerical calculations S.S. and Analytic Theory J.d.G.}



\sffamily{\textbf{A locally heated Janus colloid can achieve motion in a fluid through the coupling of dissolved ions and the medium's polarizibility to an imposed temperature gradient, an effect known as self-thermo(di)electrophoresis. We numerically study the self-propulsion of such a ``hot swimmer'' in a monovalent electrolyte solution using the finite-element method. The effect of electrostatic screening for intermediate and large Debye lengths is charted and we report on the fluid flow generated by self-thermoelectrophoresis. We obtain excellent agreement between our analytic theory and numerical calculations in the limit of high salinity, validating our approach. At low salt concentrations, we consider two analytic approaches and use Teubner's integral formalism to arrive at expressions for the speed. These expressions agree semi-quantitatively with our numerical results for conducting swimmers, providing further validation. Our numerical approach provides a solid framework against the strengths and weaknesses of analytic theory can be appreciated and which should benefit the realization and analysis of further experiments on hot swimming.}}


\rmfamily 


\section{\label{sec:intro}Introduction}

Nearly a decade and a half ago saw the introduction of the first man-made chemical swimmers, colloidal particles that used catalytic decomposition of hydrogen peroxide (\ce{H2O2}) to achieve self-propulsion~\cite{paxton2004catalytic, howse2007self}. These Janus swimmers where heralded as artificial model systems for studying the complex motion and cooperative behavior observed in biology~\cite{marchetti2013hydrodynamics}; such dynamics have by now indeed been reproduced in man-made systems~\cite{theurkauff2012dynamic, palacci2013living, buttinoni2013dynamical, duan2013transition, singh2017non}. Nevertheless, despite the success of these chemical swimmers, many open problems remain regarding their application. 

For example, \ce{H2O2} is detrimental to biological systems, as are many other catalytic fuels~\cite{gao2014catalytic, zhou2018photochemically}, which limits their potential for \textit{in vivo} use. This has led to the exploration of other self-propulsion strategies, which involve biocompatible surface chemistry~\cite{dey2015micromotors, ma2016motion}. A promising alternative to chemical self-propulsion is thermophoresis~\cite{jiang2010active, buttinoni2012active, baraban2013fuel, simoncelli2016combined}, which utilizes local heating to achieve motion through the migration of solute species in a temperature gradient. The underlying Soret effect can make use of solutes already present in the local environment and does not require large temperature gradients; it may therefore be compatible with living systems. 

From a theoretical perspective, there remain open questions concerning the microscopic origins of the thermophoretic effect and associated Soret coefficients~\cite{piazza2008thermophoresis}, seeing very recent attempts to unify thermophoretic theory for colloidal motion~\cite{burelbach2018unified}. Significant progress has, however, been made theoretically for a specific thermal driving mechanism, where the dominant contribution comes from electrostatic interactions,~\textit{i.e.}, thermoelectrophoresis~\cite{majee2012charging, simoncelli2016combined, dietzel2016thermoelectricity, dietzel2017flow, ly2018nanoscale, burelbach2018determining, burelbach2019particle}. In addition, thermophoretic theory has been used to clarify experimental results,~\textit{e.g.}, see Refs.~\citenum{samin2015self, simoncelli2016combined}. Yet most thermo(di)electrophoretic theory considers the limit of high ionic strength, for which several analytic methods can be utilized.

In this paper, we theoretically study self-propulsion \textit{via} thermo(di)electrophoresis, for which we go beyond the thin-screening-layer (Smoluchowski-limit) approximations made in previous works~\cite{majee2012charging, simoncelli2016combined, ly2018nanoscale}. We describe in detail the associated equation system and solve it using the finite-element method (FEM) over the full range of experimentally relevant ion concentrations, from the Smoluchowski to the H{\"u}ckel (low-salt) limit. Our calculations show motility reversals that are reminiscent of those found in external electrophoresis~\cite{obrien1978electrophoretic} and those recently reported for external thermodielectrophoresis~\cite{burelbach2018determining}. 

We obtain self-propulsion speeds of a few $\si{\micro\meter\per\second}$ for physiologically relevant salt (monovalent ions) concentrations $\approx \SI{1}{\milli\mole\per\liter}$ and small local heating of $\Delta T \lesssim \SI{5}{\kelvin}$, in agreement with the literature. Changing the ions and bulk salt concentration also allows for sensitive tuning of the speed and flow field around the hot swimmer by controlling the Seebeck effect~\cite{bickel2013flow}, for which we explore the impact of low ionic strength. We complement our FEM results for the swim speed with analytic theory. This is based on an expansion both in terms of small temperature gradients, and in terms of small (gradients of) ion concentrations and potentials. We show that full linearization in both expansions is \textit{not} possible and that cross terms between equilibrium and out-of-equilibrium expansion fields must be preserved in order to account for thermo(di)electrophoretic self-propulsion.

We determine the self-propulsion speed both in the slip-layer approximation and using an integral formalism based on reciprocality, originally developed by Teubner~\cite{teubner1982motion}. This allows for direct evaluation of the swim speed from the body-force distribution, without placing constraints on the size of the screening layer with respect to the colloid. We obtain agreement between both approaches in the Smoluchowski limit. The generality of Teubner's formalism also allows us to tackle the regime of intermediate ionic strength and the H{\"u}ckel limit. We discuss two analytic approaches for studying the departures from the H{\"u}ckel limit. The first is based on regime splitting, while the second uses an \textit{ansatz} for the temperature distribution around the swimmer. The second approach is more accurate, but also more involved. For an equipotential swimmer, we obtain good agreement between our numerical and analytic speed expressions over the full range of ionic strengths considered. Our expressions for the speed fortuitously hold even when the concentration profiles and charge excess are no longer captured by the analytic theory. For an insulating swimmer, only the Smoluchowski limit is well captured.

The remainder of this manuscript is structured as follows. In~\cref{sec:model} we introduce the model system. \Cref{sec:theory} provides the linearization of the equation system that governs the self-thermo(di)electrophoretic motion. \Cref{sec:thin} obtains swim speeds from these linearized expressions in the Smoluchowski limit. \Cref{sec:thick} details our analytic calculations based on Teubner's formalism around the H{\"u}ckel limit. Lastly, \cref{sec:intermediate} provides an analysis of the speed for intermediate ionic strengths. The analytic sections are lengthy and algebra heavy and readers primarily interested in the numerical results are therefore recommended to skip ahead to~\cref{sec:result}. We conclude and provide an outlook in~\cref{sec:conclusion}.

\section{\label{sec:model}The Model}

\begin{figure}[!htb]
  \centering
  \includegraphics[width=0.9\columnwidth]{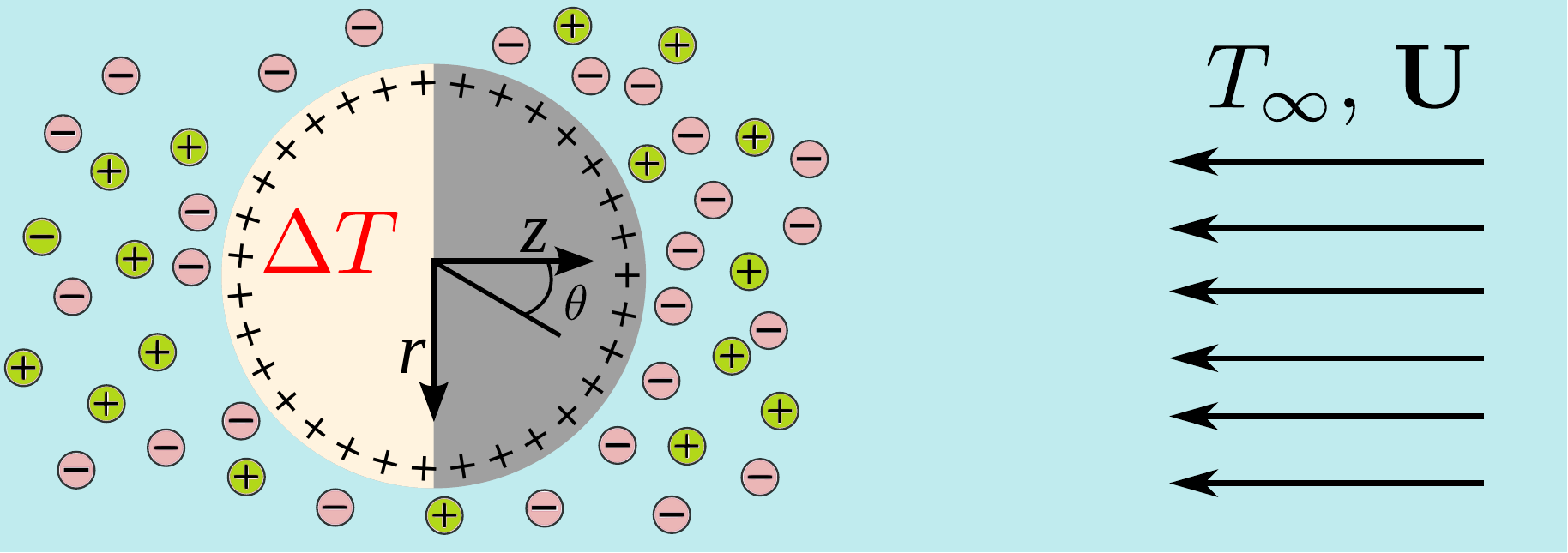}
  \caption{\label{fig:sys}Sketch of a charged Janus particle (axisymmetric around $z$) immersed in an electrolyte with an ambient temperature $T_{\infty}$. Illumination of the capped hemisphere (light yellow) increases its temperature by $\Delta T$. In steady state, the heating leads to an asymmetric distribution of ions around the colloid, resulting in its self-propulsion. In the co-moving reference frame, the fluid velocity is then $\vec{U}$ at infinity.}
\end{figure}

We consider a single spherical colloid of radius $a$ with its bottom half coated by a thin metal or carbon cap. The colloid is immersed in an electrolyte, comprised of water and a monovalent salt, with reservoir concentration $n^{\infty}$ and local salinities $n_{\pm}(\vec{r})$, where $\vec{r}$ is the position vector. By illuminating the colloid with an appropriately chosen light source, the cap can be heated, which leads to a temperature heterogeneity around the colloid which drives the system out of equilibrium. This causes the colloid to self-propel due to the thermoelectrophoresis, see~\cref{fig:sys} for a schematic illustration. Here, we also define our radial $r$ and axial $z$ coordinates (unit vectors $\hat{r}$ and $\hat{z}$, respectively), as well as the polar angle $\theta$.

The governing equations of our system in steady state are as follows. The temperature distribution throughout the system is given by $T(\vec{r})$ and obeys the heat equation,
\begin{align}
\label{eq:heat} \vec{\nabla} \cdot \left( k(T(\vec{r})) \vec{\nabla} T(\vec{r}) \right) &= 0~,
\end{align}
where $k$ is the thermal conductivity, with $k=k_{\mathrm{f}}$ in the fluid and $k=k_{\mathrm{s}}$ for the solid colloid. In~\cref{eq:heat}, we neglected advection in the fluid phase since the typical $\mathcal{O}(\si{\micro\meter\per\second})$ velocities of microswimmers lead to small thermal P{\'e}clet numbers. Note that we take the thermal conductivity in~\cref{eq:heat} to be temperature dependent, with the constitutive relation for $k(T)$ given in~\cref{sec:theory}. Temperature dependence will be considered for all physical properties in this work. However, we leave the $T(\vec{r})$ dependence of all fields,~\textit{e.g.}, the fluid velocity and potential, implicit throughout. 

Within a continuum framework, the ion dynamics is captured by the classical Poisson-Nernst-Planck equations. The Poisson equation for the electric potential $\Phi(\vec{r})$ reads
\begin{align}
\label{eq:Poisson} \vec{\nabla} \cdot \left( \epsilon(T(\vec{r})) \vec{\nabla} \Phi(\vec{r}) \right) &= - e \left( n_{+}(\vec{r}) - n_{-}(\vec{r}) \right)~,
\end{align}
where $\epsilon$ is the medium's dielectric permittivity and $e$ is the elementary charge. The Nernst-Planck equations for the ion fluxes are~\cite{majee2012charging, dietzel2017flow}
\begin{align}
\nonumber \vec{j}_{\pm}(\vec{r}) &= -D_{\pm}(T(\vec{r})) \bigg[ \vec{\nabla} n_{\pm}(\vec{r}) \pm \frac{ e n_{\pm}(\vec{r})}{k_{\mathrm{B}} T(\vec{r})} \vec{\nabla} \Phi(\vec{r}) \\
\label{eq:flux} & \phantom{= -D_{\pm}(T(\vec{r})) \bigg[} + 2 n_{\pm}(\vec{r}) \alpha_{\pm}(T(\vec{r})) \frac{\vec{\nabla} T(\vec{r})}{T(\vec{r})} \bigg]~,
\end{align}
where $k_{\mathrm{B}}$ is Boltzmann's constant, $D_{\pm}$ are the regular diffusion constants, and $\alpha_{\pm}$ are the thermal diffusivities of the respective ions. The latter are related to the intrinsic Soret coefficients \textit{via} $S_{\pm} = 2 \alpha_{\pm}/T$. The equation system is closed by the ionic conservation laws
\begin{align}
\label{eq:cons} \vec{\nabla} \cdot \vec{j}_{\pm}(\vec{r}) &= 0~,
\end{align}
where we have employed the low-P{\'e}clet-number approximation to eliminate advective ion transport. That is, ionic diffusion dominates advection, see~\cref{app:justify} for the justification.

For a micron-size colloid self-propelling in water at a speed that is $\mathcal{O}\left(\SI{1}{\micro\meter\per\second}\right)$, the relevant Reynolds number $\mathrm{Re} \ll 1$. The fluid velocity is thus governed by the Stokes equations for an incompressible fluid
\begin{align}
\nonumber \eta(T(\vec{r})) \underline{\Delta} \vec{u}(\vec{r}) - \vec{\nabla} p(\vec{r}) &= e \left( n_{+}(\vec{r}) - n_{-}(\vec{r}) \right) \vec{\nabla} \Phi(\vec{r}) \\
\label{eq:stokes} &\phantom{=} + \frac{1}{2} \left\vert \vec{\nabla} \Phi(\vec{r}) \right\vert^{2} \frac{\partial \epsilon(T(\vec{r}))}{\partial T(\vec{r})} \vec{\nabla} T(\vec{r}) ~; \\
\label{eq:mass} \vec{\nabla} \cdot \vec{u}(\vec{r}) &= 0~,
\end{align}
where $\eta$ is the viscosity of the solvent and $p(\vec{r})$ is the hydrostatic pressure; $\underline{\Delta}$ indicates the vector Laplacian. Here, we use in the right-hand side of~\cref{eq:stokes} the body-force terms derived by Landau and L{\'i}fsh{\'i}ts~\cite{landau1984electrodynamics} and also employed by Refs.~\citenum{morozov1999thermal, rasuli2008soret, wurger2008transport, ly2018nanoscale}, where the first term is the electric body force, which implicitly depends on the temperature through the ionic distributions, and the second term is the thermoelectric coupling due to the permittivity dependence on temperature. 

The boundary conditions for our problem are the following. On the swimmer, we have a no-slip condition for the fluid velocity, $\vec{u}(\vec{r}_{s}) = \vec{0}$, where $\vec{r}_{s}$ is a position vector on the surface of the swimmer; $\vert \vec{r}_{s} \vert = a$. We choose a frame of reference co-moving with the particle such that, the fluid velocity far away from the particle obeys $\vec{u}(\vert \vec{r} \vert \uparrow \infty) = -\vec{U}$, with $\vec{U}$ the swim velocity and $U = \vec{U} \cdot \hat{z}$ the swim speed. N.B. Our definition of the swim speed allows it to assume negative values, which we use throughout to help identify the direction of travel. 

The Poisson equation has the boundary condition that the electrostatic potential decays to zero in the bulk,~\textit{i.e.}, $\Phi(\vert \vec{r} \vert \uparrow \infty) = 0$. At the surface, we must distinguish between a conductor and an insulator. For the former, we have $\Phi(\vec{r}_{s}) = \Phi_{0}(\vec{r}_{s})$, with $\Phi_{0}$ the surface potential. For the latter, we have $\hat{n}(\vec{r}_{s}) \cdot \left. \vec{\nabla} \Phi(\vec{r}) \right\vert_{\vec{r} = \vec{r}_{s}} = - \sigma(\vec{r}_{s})/\epsilon(T(\vec{r}_{s}))$, where $\sigma$ is the surface charge density and $\hat{n}$ is the outward unit normal to the surface. The salt concentrations at the edge of the system assume their reservoir value, $n_{\pm}(\vert \vec{r} \vert \uparrow \infty) = n^{\infty}$. At the surface, we employ no penetration boundary conditions for the ionic species: $\hat{n}(\vec{r}_{s}) \cdot \vec{j}_{\pm}(\vec{r}_{s}) = 0$. 

Finally, for the heat equation, the temperature far away is given by the reservoir temperature $T(\vert \vec{r} \vert \uparrow \infty) = T^{\infty}$. For the capped surface we must distinguish between constant heat flux and constant temperature, respectively. When the thermal conductivity of the coating $k_{\mathrm{cap}}$ is much larger than that of the fluid and solid colloid, $k_{\mathrm{f}}$ and $k_{\mathrm{s}}$, respectively, there is a constant temperature on the lower hemisphere $T(\vec{r}_{s}) = T^{\infty} + \Delta T$, with $\Delta T$ the excess temperature induced by heating. This typically occurs for a metallic cap~\cite{buttinoni2012active, buttinoni2013dynamical}. When thermal conductivity of the coating is much smaller that of the fluid and colloid, $k_{\mathrm{cap}} \ll k_{\mathrm{f}},~k_{\mathrm{s}}$,~\textit{e.g.}, for a carbon coating~\cite{gomez2017tuning}, heat is immediately conducted to the surroundings such that the illumination leads to a constant heat flux $Q$ through the cap. In this case, the boundary condition reads $k_{\mathrm{s}} \hat{n}(\vec{r}_{s}) \cdot \left. \vec{\nabla} T(\vec{r}) \right\vert_{\vec{r} = \vec{r}_{s}} - k_{\mathrm{f}} \hat{n}(\vec{r}_{s}) \cdot \left. \vec{\nabla} T(\vec{r}) \right\vert_{\vec{r} = \vec{r}_{s}} = Q(\vec{r}_{s})$. On the top (uncapped) half of the colloid, we have the flux continuity condition $k_{\mathrm{s}} \hat{n}(\vec{r}_{s}) \cdot \left. \vec{\nabla} T(\vec{r}) \right\vert_{\vec{r} = \vec{r}_{s}} = k_{\mathrm{f}} \hat{n}(\vec{r}_{s}) \cdot \left. \vec{\nabla} T(\vec{r}) \right\vert_{\vec{r} = \vec{r}_{s}}$.

The system of~\crefrange{eq:heat}{eq:mass} with the appropriate boundary conditions was solved numerically using the finite element software \textsf{COMSOL Multiphysics} to obtain the self-propulsion speed of the particle, see~\cref{sec:result}.

\section{\label{sec:theory}Linear Analytic Theory}

To gain deeper insight into our system, we derive expressions for the speed of the thermoelectrophoretic swimmer $U$ by linearizing ~\crefrange{eq:heat}{eq:mass}. The approach we employ is similar to that of Ref.~\citenum{brown2017ionic}, but applied here also to the temperature dependencies. Linearization is not required, however, to establish the temperature profile. This can be obtained using the expansion presented in Ref.~\citenum{bickel2013flow}; for completeness we provide the relevant expressions in our notation in~\cref{app:temp}. 

Our first linearization is of the electrostatic potential and the ion distributions,~\textit{i.e.}, we make the usual Debye-H{\"u}ckel approximation. We write 
\begin{align}
\label{eq:ionlin} n_{\pm}(\vec{r}) &= n^{\infty} \left( 1 + x_{\pm}(\vec{r}) \right)~, \\
\label{eq:potlin} \Phi(\vec{r}) &= \frac{k_{\mathrm{B}} T^{\infty}}{e} \phi(\vec{r})~,
\end{align}
where $x_{\pm}(\vec{r})$ and $\phi(\vec{r})$ are the dimensionless, linearized ion distributions and potential, respectively. 

It should be noted that in some cases, particularly those involving thermocharging~\cite{wurger2008transport, majee2012charging, ly2018nanoscale}, resorting to the Debye-H{\"u}ckel approximation is unnecessary. However, in the case of self- and external thermo(di)electrophoresis, the radial symmetry breaking due to the temperature gradient leads to complicated differential equations that only have closed-form solutions in certain cases. This assumption will turn out to limit the applicability of our analytic expressions in the H{\"u}ckel limit for insulating swimmers, see~\cref{sec:result}.

Our second linearization decomposes the fields and physical quantities into equilibrium (``eq'') and non-equilibrium (``neq'') parts, where the non-equilibrium parts are due to variations in temperature. Here, we shall expand in the small parameter $\tau \equiv \Delta T / T^{\infty}$, corresponding to the relative maximum temperature difference $\Delta T$ from the reservoir temperature $T^{\infty}$. Note that $\tau$ is well-defined for both equi-temperature and equi-flux surfaces. 

This choice of expansion parameter allows us to write to the temperature distribution as
\begin{align}
\label{eq:tnonlin}
 T(\vec{r}) = T^{\infty}\left( 1 + \tau t(\vec{r}) \right)~,
\end{align}
where $t(\vec{r})$ is the dimensionless temperature. Similarly, for the other physical fields the decomposition yields: $x_{\pm}(\vec{r}) = x_{\pm}^{\mathrm{eq}}(\vec{r}) + \tau x_{\pm}^{\mathrm{neq}}(\vec{r})$, $\phi(\vec{r}) = \phi^{\mathrm{eq}}(\vec{r}) + \tau \phi^{\mathrm{neq}}(\vec{r})$, $\vec{u}(\vec{r}) = \tau \vec{v}(\vec{r})$, and $p(\vec{r}) = p^{\mathrm{eq}}(\vec{r}) + \tau p^{\mathrm{neq}}(\vec{r})$. Notice that in equilibrium there is no fluid flow, hence we only have the out-of-equilibrium $\vec{v}$ velocity component. We simplify the equations further by introducing the conjugate variables to the ionic distributions: the local salinity, $X(\vec{r}) \equiv \left( x_{+}(\vec{r}) + x_{-}(\vec{r}) \right)/2$, and the local ion excess or space charge density, $\delta \! X(\vec{r}) \equiv \left( x_{+}(\vec{r}) - x_{-}(\vec{r}) \right)/2$.

The physical quantities are expanded as: $\epsilon/\epsilon^{\infty} = 1 + \tau \epsilon^{\ast} t(\vec{r})$, $\eta/\eta^{\infty} = 1 + \tau \eta^{\ast} t(\vec{r}) $, $D_{\pm}/ D_{\pm}^{\infty} = 1 + \tau D_{\pm}^{\ast} t(\vec{r})$, $\alpha_{\pm}/ \alpha_{\pm}^{\infty} = 1 + \tau \alpha_{\pm}^{\ast} t(\vec{r}) $, $k/k^{\infty} = 1 + \tau k^{\ast} t(\vec{r})$. Here, the ``$\infty$'' superscript denotes the reservoir value, which is located at infinity, and the ``$\ast$'' superscript the first-order Taylor expansion coefficient. We have numerically verified that all starred quantities are order unity and that the non-equilibrium fields are much smaller than the equilibrium contributions, see~\cref{app:justify}. 

We now use the above perturbative expressions to expand all equations in terms of $\tau$, keeping only the zeroth-order and first-order terms. The zeroth order gives the equilibrium equations at constant temperature $T^{\infty}$, which are the standard linear Poisson-Boltzmann equations, see~\cref{app:linearized}. The solution of the linear equilibrium problem is $x_{\pm}^{\mathrm{eq}}(\vec{r}) = \mp \phi^{\mathrm{eq}}(\vec{r})$. The first-order equations capture the leading out-of-equilibrium effects. 

The heat equation becomes $\nabla^{2} t(\vec{r}) = 0$, for which the solution is given in~\cref{app:temp}. The Poisson equation reduces to
\begin{align}
\label{eq:Poissonlinneqred} \nabla^{2} \phi^{\mathrm{neq}}(\vec{r}) + \epsilon^{\ast} \vec{\nabla} \cdot \left( t(\vec{r}) \vec{\nabla} \phi^{\mathrm{eq}}(\vec{r}) \right) &= - (\kappa^{\infty})^{2} \delta \! X^{\mathrm{neq}}(\vec{r})~,
\end{align}
with the inverse reservoir Debye length
\begin{align}
\label{eq:kappa} \kappa^{\infty} &\equiv \sqrt{\frac{2 e^{2} n^{\infty}}{\epsilon^{\infty} k_{\mathrm{B}} T^{\infty}}}~.
\end{align}
The Stokes equations read
\begin{align}
\nonumber \eta^{\infty} \underline{\Delta} \vec{v}(\vec{r}) - \vec{\nabla} p^{\mathrm{neq}}(\vec{r}) &= \\
\nonumber 2 k_{\mathrm{B}} T^{\infty} n^{\infty} \Bigg[ \delta \! X^{\mathrm{neq}}(\vec{r}) \vec{\nabla} \phi^{\mathrm{eq}}(\vec{r}) - \phi^{\mathrm{eq}}(\vec{r}) \vec{\nabla} \phi^{\mathrm{neq}}(\vec{r}) & \\
\label{eq:stokeslinneqred} + \frac{1}{2} \left( \lambda^{\infty} \right)^{2} \epsilon^{\ast} \left\vert \vec{\nabla} \phi^{\mathrm{eq}}(\vec{r}) \right\vert^{2} \vec{\nabla} t(\vec{r}) \Bigg]~, & \\
\label{eq:massconsred} \vec{\nabla} \cdot \vec{v}(\vec{r}) &= 0~, 
\end{align}
 where we used $\partial \epsilon / \partial T = \epsilon^{\infty} \epsilon^{\ast} / T^{\infty}$ and introduced the Debye length $\lambda^{\infty} \equiv 1/ \kappa^{\infty}$. Finally, the ionic fluxes become
\begin{align}
\nonumber \vec{j}_{\pm}^{\mathrm{neq}}(\vec{r}) &= - D_{\pm}^{\infty} n^{\infty} \big[ \vec{\nabla} x_{\pm}^{\mathrm{neq}}(\vec{r}) \pm \vec{\nabla} \phi^{\mathrm{neq}}(\vec{r}) + 2 \left( 1 \mp \phi^{\mathrm{eq}}(\vec{r}) \right) \alpha_{\pm}^{\infty} \vec{\nabla} t(\vec{r})\\
\label{eq:fluxlinneq} & \phantom{= - D_{\pm}^{\infty} n^{\infty} \big[} \mp t(\vec{r}) \vec{\nabla} \phi^{\mathrm{eq}}(\vec{r}) \big] ~, \\
\label{eq:conslinneq} \vec{\nabla} \cdot \vec{j}_{\pm}^{\mathrm{neq}}(\vec{r}) &= 0~,
\end{align}
Note that at linear order in $\tau$, only the term involving $\epsilon^{\ast}$ in~\cref{eq:Poissonlinneqred} introduces the temperature dependence of the physical properties into the equations. This is why a thermo\textit{di}electrophoretic component to this phoresis has been reported in the literature~\cite{morozov1999thermal, rasuli2008soret, wurger2008transport, ghonge2013electrohydrodynamics, ly2018nanoscale}. The reason why the $D_{\pm}^{\ast}$ terms drop out, is that they are paired to first order with equilibrium fluxes, which are vanishing.

By adding and subtracting the flux expressions in~\cref{eq:fluxlinneq}, and employing the conservation equations~\eqref{eq:conslinneq}, we obtain
\begin{align}
\label{eq:Xred} \nabla^{2} X^{\mathrm{neq}}(\vec{r}) &= \beta \vec{\nabla} t(\vec{r}) \cdot \vec{\nabla} \phi^{\mathrm{eq}}(\vec{r})~; \\
\nonumber \nabla^{2} \delta \! X^{\mathrm{neq}}(\vec{r}) + \nabla^{2} \phi^{\mathrm{neq}}(\vec{r}) &= \left( 1 + \gamma \right) \vec{\nabla} t(\vec{r}) \cdot \vec{\nabla} \phi^{\mathrm{eq}}(\vec{r}) \\
\label{eq:dXred} & \phantom{=} + (\kappa^{\infty})^{2} t(\vec{r}) \phi^{\mathrm{eq}}(\vec{r})~,
\end{align}
where we have used $\nabla^{2} t(\vec{r}) = 0$ and introduced
\begin{align}
\label{eq:beta} \beta &\equiv \alpha_{+}^{\infty} - \alpha_{-}^{\infty}~, \\
\label{eq:gamma} \gamma &\equiv \alpha_{+}^{\infty} + \alpha_{-}^{\infty}~,
\end{align}
with $\beta$ commonly referred to as the (reduced) Seebeck parameter. The expressions for the non-equilibrium flux contain two temperature-related terms. One coupling the gradient of the temperature to the equilibrium ion distributions \textit{via} the thermal diffusion constant. The other coupling the temperature itself to the gradient of the equilibrium electrostatic potential, \textit{via} the ionic mobilities~\cite{dietzel2016thermoelectricity, dietzel2017flow}. The latter is represented by the term on the second line of~\cref{eq:fluxlinneq}. Ultimately, this implies that there can be thermoelectrophoretic swimming without any thermal-diffusion ($\alpha_{\pm}$-related) effect, also see Ref.~\citenum{ly2018nanoscale}. From~\cref{eq:dXred} it follows that a non-equilibrium ionic excess will be present even if the thermal diffusion coefficients are zero, in agreement with Refs.~\citenum{dietzel2016thermoelectricity, dietzel2017flow, ly2018nanoscale}. 

Establishing a solution to the above equation system is nontrivial. Equations~\eqref{eq:Xred} and~\eqref{eq:dXred} reveal that the cross coupling between temperature fields and equilibrium ionic screening is crucial to obtain thermoelectrophoresis. If we ignore such cross terms, only the trivial solution is obtained. This intrinsic nonlinearity complicates obtaining solutions using standard spectral methods~\footnote{By ``nonlinearity'' we mean here that the differential equation system cannot be cast into the `standard form' of a Laplacian acting on a vector comprising the individual expansion fields equated to a coefficient matrix acting on the same vector, as was,~\textit{e.g.}, done in Ref.~\citenum{brown2017ionic}. This hinders a solution strategy based on orthogonalization of this matrix and recovery of the relevant decay lengths as the diagonal elements of the resulting eigenmatrix.}. Nevertheless, we stress that the theory is still fully linear in terms of the temperature dependence. 

Further note that the fields $\delta \! X^{\mathrm{neq}}(\vec{r})$ and $\phi^{\mathrm{neq}}(\vec{r})$ form a closed subsystem of equations, to linear order in $\tau$. The non-equilibrium ion concentration $X^{\mathrm{neq}}$ in~\cref{eq:Xred} is only due to coupling between the temperature gradient and the equilibrium ion potential. It is weighted by the difference in thermal diffusivity $\beta$, meaning that $X^{\mathrm{neq}}$ vanishes, when there is no thermal-diffusion-based ion accumulation in the double layer ($\alpha_{+}^{\infty} = \alpha_{-}^{\infty}=0$ or $\alpha_{+}^{\infty} = \alpha_{-}^{\infty}$). The closed subsystem is the only non-equilibrium part that then remains. This feature suggests a route toward solving the full set of equations, which is explored in~\cref{sec:intermediate}.

\section{\label{sec:thin}The Smoluchowski Limit}

In this section, we will limit ourselves to the case of high ionic strength and make the thin-screening-layer approximation. We compute the swim speed both using the slip-layer approximation and Teubner's integral formalism~\cite{teubner1982motion} to double check the expressions. In~\cref{sec:result} we validate these further using our FEM calculations.

\subsection{\label{sub:ele}The Electrostatic Potential and Ion Profiles}

In the high-ionic-strength or Smoluchowski limit, the electrostatic screening length $\lambda^{\infty}$ is small compared to the particle radius, $\kappa^{\infty} a \gg 1$. Outside (``out'') of the screening layer, we have $\nabla^{2} X^{\mathrm{neq}}_{\mathrm{out}}(\vec{r}) = 0$ and $\nabla^{2} \phi^{\mathrm{neq}}_{\mathrm{out}}(\vec{r}) = 0$, since $\phi^{\mathrm{eq}}(\vec{r}) = 0$ in this region and $\delta \! X^{\mathrm{neq}}_{\mathrm{out}}(\vec{r}) = 0$, because any excess charge is screened. This implies that the only solutions for the potential and total salinity permissible in the region outside of the double layer have a Laplace form. We know that the temperature satisfies this equation and that it sets up fluxes of ions in the bulk. These fluxes will thus be proportional to $t$. 

Clearly, in the bulk there cannot be charge separation due to differences in thermal diffusivity, otherwise a net charge would appear. Hence, an \textit{unscreened} non-equilibrium potential is set up to prevent this, which is proportional to $\beta$. Furthermore, ion transport in a thermal gradient can effect the local salinity, as ions may be repelled or drawn towards areas of higher temperature; this effect will scale with $\gamma$. Using~\cref{eq:fluxlinneq}, one indeed finds that $X^{\mathrm{neq}}_{\mathrm{out}}(\vec{r}) = -\gamma t(\vec{r})$ and $\phi^{\mathrm{neq}}_{\mathrm{out}}(\vec{r}) = -\beta t(\vec{r})$, in agreement Ref.~\citenum{majee2012charging}. 

Now let $\lambda^{\infty} q$ measure distance in the direction perpendicular to the surface, with $q = 0$ for $r = a$. In our approximation, the curvature of the sphere can locally be ignored. We can then split the solutions into parallel and perpendicular components: $\phi^{\mathrm{eq}}(\vec{r}_{s},q) = \phi(\vec{r}_{s}) e^{-q}$ and $t(\vec{r}_{s},q) = t(\vec{r}_{s})$. Here, $\phi(\vec{r}_{s})$ is the potential at the surface and $t(\vec{r}_{s})$ is the temperature at the surface. On the size of the screening layer, the temperature is approximately radially constant because the temperature in the fluid decays with a power law of leading order $a/r$. 

In this limit, the boundary conditions at the surface of the particle ($q=0$) need to be determined. N.B. Here, we do not consider temperature-dependent charge regulation. We linearize the conducting and insulating conditions, leading to $\phi^{\mathrm{neq}}_{\mathrm{in}}(\vec{r}_{s},0) = 0$ and $\left. \partial_{q} \phi^{\mathrm{neq}}_{\mathrm{in}}(\vec{r}_{s},q) \right\vert_{q = 0} = 0 $, respectively. For a \textit{conductor}, the equilibrium part of the field accounts fully for the surface potential, $\phi^{\mathrm{eq}}_{\mathrm{in}}(\vec{r}_{s},0) = \phi_{0}(\vec{r}_{s})$, with $\phi_{0}(\vec{r}_{s})$ the reduced surface potential. This implies $\phi^{\mathrm{eq}}_{\mathrm{in}}(\vec{r}_{s},q) = \phi_{0}(\vec{r}_{s}) e^{-q}$. For an \textit{insulator}, the boundary condition of the equilibrium potential
\begin{align}
\label{eq:surfchar} \left. \partial_{q} \phi^{\mathrm{eq}}_{\mathrm{in}}(\vec{r}_{s},q) \right\vert_{q = 0} &= - \frac{\lambda^{\infty} e \sigma(\vec{r}_{s}) }{ \epsilon^{\infty} k_{\mathrm{B}} T^{\infty} }~,
\end{align}
covers any surface charge present. This implies that $\phi^{\mathrm{eq}}_{\mathrm{in}}(\vec{r}_{s},q) = \lambda^{\infty} e \sigma(\vec{r}_{s}) e^{-q} /( \epsilon^{\infty} k_{\mathrm{B}} T^{\infty} )$, which matches a series expansion of the full solution for the curved surface in terms of $\lambda^{\infty} \ll a$.

We can thus generally write $\phi^{\mathrm{eq}}_{\mathrm{in}}(\vec{r}_{s},q) = \phi(\vec{r}_{s}) e^{-q}$, with $\phi(\vec{r}_{s})$ either $\phi_{0}(\vec{r}_{s})$ (conducting) or 
$ \lambda^{\infty} e \sigma(\vec{r}_{s})/( \epsilon^{\infty} k_{\mathrm{B}} T^{\infty} )$ (insulating). The fact that the insulating case has a prefactor $\lambda^{\infty}$ does not have consequences for the expansions that will be performed next, as the parallel components scale $\mathcal{O} (1)$, and all other components either $\mathcal{O} ( \kappa^{\infty} )$ or 
$\mathcal{O} \left( (\kappa^{\infty})^{2} \right)$. However, as we will see in~\cref{sub:potvssurf}, it will have consequences for the speeds that can be achieved in the Smoluchowski limit by an insulating swimmer.

Applying the coordinate transformation inside the screening layer, we have for the Laplacian: $\nabla^{2} = \nabla_{\parallel}^{2} + \left( \kappa^{\infty} \right)^2 \partial_{q}^{2}$, with $\vec{\nabla}_{\parallel}$ the gradient in the tangent plane to $\vec{r}_{s}$ and $\nabla_{\parallel}^{2}$ the associated in-plane Laplacian. Taking the limit $\lambda^{\infty} \downarrow 0$, leads to the following transformed salinity and space charge density
\begin{align}
\label{eq:Xthin} \partial_{q}^{2} X^{\mathrm{neq}}_{\mathrm{in}}(\vec{r}_{s},q) &= 0~; \\
\label{eq:dXthin} \partial_{q}^{2} \delta \! X^{\mathrm{neq}}_{\mathrm{in}}(\vec{r}_{s},q) + \partial_{q}^{2} \phi^{\mathrm{neq}}_{\mathrm{in}}(\vec{r}_{s},q) &= t(\vec{r}_{s}) \phi(\vec{r}_{s}) e^{-q}~,
\end{align}
with corresponding Poisson equation
\begin{align}
\label{eq:Poissonthin} \partial_{q}^{2} \phi^{\mathrm{neq}}_{\mathrm{in}}(\vec{r}_{s},q) &= - \epsilon^{\ast} t(\vec{r}_{s}) \phi(\vec{r}_{s}) e^{-q} - \delta \! X^{\mathrm{neq}}_{\mathrm{in}}(\vec{r}_{s},q)~,
\end{align}
where the subscript ``$\mathrm{in}$'' is used to indicate that these fields are within the thin screening layer. Since only derivatives with respect to $q$ remain in~\crefrange{eq:Xthin}{eq:Poissonthin}, they can be solved using separation of variables. 

The limit $\vert \vec{r} \vert \downarrow a$ for the solutions outside of the screening layer gives a set of boundary conditions for the solution inside. Note that by construction this corresponds to $q \uparrow \infty$. Taking this limit within the layer, we find $X^{\mathrm{neq}}_{\mathrm{in}}(\vec{r}_{s},q \uparrow \infty) = -\gamma t(\vec{r}_{s}^{+})$, $\delta \! X^{\mathrm{neq}}_{\mathrm{in}}(\vec{r}_{s},q \uparrow \infty) = 0$, and $\phi^{\mathrm{neq}}_{\mathrm{in}}(\vec{r}_{s},q \uparrow \infty) = -\beta t(\vec{r}_{s}^{+})$. Here, the value of the right-hand side is evaluated at the edge of the screening layer $\vec{r}_{s}^{+} \approx \vec{r}_{s}$, to avoid the ambiguity that arises by simultaneously demanding $\lambda^{\infty} \downarrow 0$.

The above conditions, together with the linearized~\crefrange{eq:Xthin}{eq:Poissonthin}, lead to the following solutions within the screening layer. As the temperature decays very little inside the screening layer, the added total salinity due to the heating therein is therefore $X^{\mathrm{neq}}_{\mathrm{in}}(\vec{r}_{s},q) = -\gamma t(\vec{r}_{s})$. The effect on the net salt concentration in the double layer, however, is sufficiently small that local corrections to the Debye length do not have to be accounted for, since $X_{\mathrm{in}} = X^{\mathrm{eq}}_{\mathrm{in}} + \tau X^{\mathrm{neq}}_{\mathrm{in}}$ with $\tau \ll 1$. The non-equilibrium space charge density decays with $q$, and we must consider conducting (equipotential) and insulating (fixed charge) surfaces separately.

For an \textit{equipotential surface} (or conductor), we find
\begin{align}
\label{eq:dXinEPsol}  \delta \! X^{\mathrm{neq}}_{\mathrm{in}}(\vec{r}_{s},q) &= - \beta t(\vec{r}_{s}) e^{-q} + \frac{1}{2} \phi(\vec{r}_{s}) t(\vec{r}_{s}) \left( 2 - \left( 1 + \epsilon^{\ast} \right) q \right) e^{-q} ~,
\end{align}
where the electrostatic potential is given by
\begin{align} 
\label{eq:PinEPsol} \phi^{\mathrm{neq}}_{\mathrm{in}}(\vec{r}_{s},q) &= - \beta t(\vec{r}_{s}) \left( 1 - e^{-q} \right) + \frac{1}{2} \phi(\vec{r}_{s}) t(\vec{r}_{s}) \left( 1 + \epsilon^{\ast} \right) q e^{-q}~, 
\end{align}
because $\phi^{\mathrm{neq}}_{\mathrm{in}}(\vec{r}_{s},0) = 0$. The Seebeck effect thus leads to the development of a surface thermocharge, which is given by $\delta \! X^{\mathrm{neq}}_{\mathrm{in}} = ( \phi_{0} - \beta ) t$, as follows from~\cref{eq:dXinEPsol}. This expression differs from the one found by Majee and W{\"u}rger~\cite{majee2012charging} in that we include a non-zero imposed surface potential. Moreover, by the \textit{surface} thermocharge, we mean the charge that is imposed directly at the surface, rather than the integral form that is employed in Refs.~\citenum{majee2012charging, ly2018nanoscale}, which gives the effective \textit{bulk} thermocharge built up around the swimmer due to thermophoresis. 

The charging for $\beta = 0$ follows from the right-hand side of~\cref{eq:dXthin}. Suppose for convenience that $\epsilon^{\ast} = 0$ then~\cref{eq:dXthin,eq:Poissonthin} combine to give an inhomogeneous Helmholtz equation for $\delta \! X^{\mathrm{neq}}_{\mathrm{in}}$, for which the particular solution exactly corresponds to minus the right-hand side of~\cref{eq:dXthin}. Physically, the effect is due to the difference in temperature dependence between regular diffusion and electric migration~\cite{dietzel2016thermoelectricity, dietzel2017flow}, which leads to a surface (and bulk) thermocharge on top of that induced by the Seebeck effect~\cite{majee2012charging, ly2018nanoscale}.

For a \textit{fixed surface charge}, we find
\begin{align}
\label{eq:dXinSCsol} \delta \! X^{\mathrm{neq}}_{\mathrm{in}}(\vec{r}_{s},q) &= \frac{1}{2} \phi(\vec{r}_{s}) t(\vec{r}_{s}) \left( 2 - \left( 1 + \epsilon^{\ast} \right) \left( 1 + q \right) \right) e^{-q}~,
\end{align}
where the associated electrostatic potential reads
\begin{align} 
\label{eq:PinSCsol} \phi^{\mathrm{neq}}_{\mathrm{in}}(\vec{r}_{s},q) &= - \beta t(\vec{r}_{s}) + \frac{1}{2} \phi(\vec{r}_{s}) t(\vec{r}_{s}) \left( 1 + \epsilon^{\ast} \right) \left( 1 + q \right) e^{-q}~.
\end{align}
Again, there is a thermocharging effect ($\delta \! X^{\mathrm{neq}}_{\mathrm{in}}(\vec{r}_{s},q) \ne 0$) as described in Refs.~\citenum{majee2012charging, ly2018nanoscale}. However, due to the difference in boundary conditions, there is only a non-Seebeck contribution, which means that \textit{uncharged} surfaces cannot pick up a surface thermocharge. They can pick up a bulk thermocharge, as follows from using the definition in Refs.~\citenum{majee2012charging, ly2018nanoscale}. In general, we will restrict our analysis to the case $\phi(\vec{r}_{s}) \ne 0$.

\subsection{\label{sub:force}Thermo(di)electrophoretic Force onto the Fluid}

We must first determine the forces acting on the fluid to perform the slip-layer approximation and obtain the swim speed. The equilibrium component of the force only generates a hydrostatic pressure, which cannot contribute to the generation of flow, by definition, see~\cref{app:linearized}. For simplicity and analytic convenience, we will assume that the reduced surface potential $\phi(\vec{r}_{s})$ is locally uniform,~\textit{i.e.}, $\vec{\nabla}_{\parallel} \phi(\vec{r}_{s}) = \vec{0}$. We use the general form of the Stokes equation $\eta^{\infty} \underline{\Delta} \vec{v} =  \vec{\nabla} p^{\mathrm{neq}} - \vec{f}^{\mathrm{neq}}$ to identify the out-of-equilibrium force density acting on the fluid in a first-order expansion
\begin{align}
\nonumber \frac{\vec{f}^{\mathrm{neq}}(\vec{r})}{k_{\mathrm{B}} T^{\infty} n^{\infty}} &= - 2\delta \! X^{\mathrm{neq}}(\vec{r}) \vec{\nabla} \phi^{\mathrm{eq}}(\vec{r}) + 2 \phi^{\mathrm{eq}}(\vec{r}) \vec{\nabla} \phi^{\mathrm{neq}}(\vec{r}) \\
\label{eq:fst} & \phantom{=} - \left( \lambda^{\infty} \right)^{2} \epsilon^{\ast} \left\vert \vec{\nabla} \phi^{\mathrm{eq}}(\vec{r}) \right\vert^{2} \vec{\nabla} t(\vec{r})~.
\end{align}

Outside the screening layer, the force density vanishes
\begin{align}
\label{eq:fstout} \frac{\vec{f}^{\mathrm{neq}}_{\mathrm{out}}(\vec{r})}{k_{\mathrm{B}} T^{\infty} n^{\infty}} &= \vec{0}~,
\end{align}
to linear order in $\tau$. Higher-order terms would lead to a force in the bulk, however this contribution is small, scaling with $\tau^{3}$, and is therefore ignored here. That is, there is an unscreened electric field ($\phi^{\mathrm{neq}}_{\mathrm{out}}(\vec{r}) = -\beta t(\vec{r})$) and a temperature gradient in the bulk. The latter would lead to a temperature-variation of the dielectric permittivity, while the former ensures that the potential prefactor in the permittivity term of~\cref{eq:stokes} is nonzero. Hence there will be a force contribution in the bulk even with strong screening.

Inside of the screening layer, we can split the driving forces into components parallel and perpendicular to the surface. These read
\begin{align}
\label{eq:fstinpar} \frac{\vec{f}^{\mathrm{neq}}_{\mathrm{in},\parallel}(\vec{r}_{s},q)}{k_{\mathrm{B}} T^{\infty} n^{\infty}} &= 2 \phi(\vec{r}_{s}) e^{-q} \vec{\nabla}_{\parallel} \phi^{\mathrm{neq}}(\vec{r}_{s},q) - \epsilon^{\ast} \phi(\vec{r}_{s})^{2} e^{-2q} \vec{\nabla}_{\parallel} t(\vec{r}_{s})~;\\ 
\label{eq:fstinper} \frac{\vec{f}^{\mathrm{neq}}_{\mathrm{in},\perp}(\vec{r}_{s},q)}{k_{\mathrm{B}} T^{\infty} n^{\infty}} &= 2\kappa^{\infty} \phi(\vec{r}_{s})e^{-q} \left[ \delta \! X^{\mathrm{neq}}(\vec{r}_{s},q) + \partial_{q} \phi^{\mathrm{neq}}(\vec{r}_{s},q) \right] \hat{q}~,
\end{align}
where we have used that $t(\vec{r})$ is almost constant in the perpendicular direction over the length of the double layer, as we did previously in~\cref{sub:ele}. We also introduced the unit normal direction to the sphere $\hat{q} $.

The expressions for the force density inside the screening layer can be rewritten using the expressions for the density and potential. For the perpendicular component we find
\begin{align}
\label{eq:fstinpersubspot} \frac{\vec{f}^{\mathrm{neq}}_{\mathrm{in},\perp}(\vec{r}_{s},q)}{k_{\mathrm{B}} T^{\infty} n^{\infty}} &= - \kappa^{\infty} \left[ 4 \beta - \left( 3 + \epsilon^{\ast} - 2 \left( 1 + \epsilon^{\ast} \right) q \right) \phi(\vec{r}_{s}) \right] \phi(\vec{r}_{s}) e^{-2q} t(\vec{r}_{s}) \hat{q}~,
\end{align}
which holds for the equipotential surface, and
\begin{align}
\label{eq:fstinpersubscha} \frac{\vec{f}^{\mathrm{neq}}_{\mathrm{in},\perp}(\vec{r}_{s},q)}{k_{\mathrm{B}} T^{\infty} n^{\infty}} &= \kappa^{\infty} \left[ 1 - \epsilon^{\ast} - 2 \left( 1 + \epsilon^{\ast} \right) q \right] \phi(\vec{r}_{s})^{2} e^{-2q} t(\vec{r}_{s}) \hat{q}~, 
\end{align}
which holds for insulating surface. Under the same assumption, we obtain
\begin{align}
\nonumber \frac{\vec{f}^{\mathrm{neq}}_{\mathrm{in},\parallel}(\vec{r}_{s},q)}{k_{\mathrm{B}} T^{\infty} n^{\infty}} &= - \left[ 2 \left( 1 - e^{-q} \right) \beta + \left( \epsilon^{*} - \left( 1 + \epsilon^{*} \right) q \right) \phi(\vec{r}_{s}) e^{-q} \right]\\
\label{eq:fstinparpot} & \phantom{=} \times \phi(\vec{r}_{s}) e^{-q} \vec{\nabla}_{\parallel} t(\vec{r}_{s})~; \\
\label{eq:fstinparcha} \frac{\vec{f}^{\mathrm{neq}}_{\mathrm{in},\parallel}(\vec{r}_{s},q)}{k_{\mathrm{B}} T^{\infty} n^{\infty}} &= - \left[ 2\beta - \left( 1 + \left( 1 + \epsilon^{*} \right) q \right) \phi(\vec{r}_{s}) e^{-q} \right] \phi(\vec{r}_{s}) e^{-q} \vec{\nabla}_{\parallel} t(\vec{r}_{s}) ~,
\end{align}
for the equipotential and insulating surface, respectively.

\subsection{\label{sub:slip}The Slip-Layer Approximation}

At this point of the calculation, we make the slip-layer approximation. We have already assumed a high ionic strength and therefore all the thermoelectrophoretic speed is generated in a small layer around the colloid. This implies that no-slip boundary condition on the colloid may be replaced by an effective slip/velocity boundary that accounts for the speed generation in the thin Debye layer,~\textit{e.g.}, see Refs.~\citenum{derjaguin1941thermo, anderson1989colloid, brady2011particle}.

Decomposing the Stokes equations into parallel and perpendicular components, we obtain the following expressions:
\begin{align}
\label{eq:stokespara} \eta^{\infty} \left( \kappa^{\infty} \right)^{2} \partial_{q}^{2} \vec{v}_{\parallel}(\vec{r}_{s},q) &=  \vec{\nabla}_{\parallel} p^{\mathrm{neq}}(\vec{r}_{s},q) - \vec{f}^{\mathrm{neq}}_{\mathrm{in},\parallel}(\vec{r}_{s},q)~; \\
\label{eq:stokesperp} \kappa^{\infty} \hat{q} \partial_{q} p^{\mathrm{neq}}(\vec{r}_{s},q) &= \vec{f}^{\mathrm{neq}}_{\mathrm{in},\perp}(\vec{r}_{s},q)~,
\end{align}
where in~\cref{eq:stokespara} the decomposed Laplacian acts only on the parallel velocity components and only the double derivative with respect to $q$ remains; it dominates due to the $\left( \kappa^{\infty} \right)^{2}$ prefactor. In~\cref{eq:stokesperp}, we used that the perpendicular fluid velocity (\textit{i.e.}, toward the particle) in the thin layer must be zero, due to incompressibility. Solving for the $q$-dependence of the pressure using~\cref{eq:stokesperp}, we find the following expressions
\begin{align}
\label{eq:pneqp} \frac{p^{\mathrm{neq}}(\vec{r}_{s},q)}{ k_{\mathrm{B}} T^{\infty} n^{\infty} } &=  \left[ 2 \beta - \left( 1 - \left( 1 + \epsilon^{\ast} \right) q \right) \phi(\vec{r}_{s}) \right] \phi(\vec{r}_{s}) t(\vec{r}_{s}) e^{-2q}~ ; \\
\label{eq:pneqc} \frac{p^{\mathrm{neq}}(\vec{r}_{s},q)}{ k_{\mathrm{B}} T^{\infty} n^{\infty} } &= \left[ \epsilon^{\ast} + \left( 1 - \left( 1 + \epsilon^{\ast} \right) q \right) \right] \phi(\vec{r}_{s})^{2} t(\vec{r}_{s}) e^{-2q}~,
\end{align}
for conducting and insulating surfaces, respectively. Here, we have set the non-equilibrium pressure to zero for ($q \uparrow 0$), because we already subtracted the equilibrium component in our linearization. 

One can group the parallel pressure gradient and parallel force terms in~\cref{eq:stokespara} and solve the resulting differential equation for $\vec{v}_{\parallel}(\vec{r}_{s},q)$, with boundary condition $\vec{v}_{\parallel}(\vec{r}_{s},q \downarrow 0) = \vec{0}$. The slip speed may be obtained by taking the limit $\vec{v}_{\mathrm{slip}}(\vec{r}_{s}) = \lim_{q \rightarrow \infty} \vec{v}_{\parallel}(\vec{r}_{s},q)$ and it is given by
\begin{align}
\label{eq:vslip} \vec{v}_{\mathrm{slip}}(\vec{r}_{s}) &= \frac{\left( \lambda^{\infty} \right)^{2}}{4 \eta^{\infty}} k_{\mathrm{B}} T^{\infty} n^{\infty} \left( 8 \beta - (1 - \epsilon^{\ast}) \phi(\vec{r}_{s})  \right) \phi(\vec{r}_{s}) \vec{\nabla}_{\parallel} t(\vec{r}_{s}) ~.
\end{align}
The above result is quite surprising, as the slip velocity has the same functional form for both insulating and conducting surfaces, also see Ref.~\citenum{ly2018nanoscale}.

The speed of the particle is now obtained by evaluating the integral
\begin{align}
\label{eq:slipvel} \bar{U} &= -\frac{1}{4 \pi a^{2}} \oint \vec{v}_{\mathrm{slip}} \mathrm{d} \vec{r}_{s}~.
\end{align}
where $\vec{v}_{\mathrm{slip}}$ is the slip velocity and integration takes place over the particle's surface. We obtain for the total speed of a swimmer
\begin{align}
\label{eq:speed_result} \bar{U} &= -\frac{k_{\mathrm{B}} T^{\infty} n^{\infty}}{6 \eta^{\infty} a} \left( \lambda^{\infty} \right)^{2} \left[ 8 \beta \phi(\vec{r}_{s}) - \left( 1 - \epsilon^{*} \right) \phi(\vec{r}_{s})^{2} \right] \bar{t}_{1}~, 
\end{align}
to leading order in $\lambda^{\infty}$. Here, $\bar{t}_{1}$ is the first Legendre-Fourier coefficient in a decomposition of the temperature, see~\cref{app:temp}.

\subsection{\label{sub:speed}The Speed according to Teubner}

We verify the slip-layer swim speed by employing Teubner's method~\cite{teubner1982motion} of integrating the (out-of-equilibrium) body force density with an integration kernel $\underline{K}(\vec{r})$ to obtain the reduced swim speed
\begin{align}
\label{eq:Tspeed} \bar{U} &= \frac{1}{6 \pi \eta^{\infty} a} \int_{V} \underline{K}(\vec{r}) \cdot \vec{f}^{\mathrm{neq}}(\vec{r}) \mathrm{d} \vec{r} ; \\
\label{eq:ker} \underline{K}(\vec{r}) &= \left( \frac{3a}{2r} - \frac{a^{3}}{2r^{3}} - 1 \right) \cos \theta \hat{r} - \left( \frac{3a}{4r} + \frac{a^{3}}{4r^{3}} - 1 \right) \sin \theta \hat{\theta}~,
\end{align}
where integration takes place over the volume $V$ outside of the particle. This integration kernel isolates \textit{via} projection those components of the body force that contribute to the speed of the particle, from those that create higher-order (multipolar) flow fields that do not contribute to locomotion. That is, the entries in the kernel have the same decay as found for fluid velocity around a dragged sphere. Note that our integral is thus also reminiscent some of the steps taken in the reciprocal approach recently proposed by Burelbach and Stark~\cite{burelbach2018determining}. 

The integral in~\cref{eq:ker} can be split into a part inside and outside of the screening layer. Then, an expansion in terms of $\lambda^{\infty}$ is performed on the integration kernel $\underline{K}(\vec{r})$ for the inner part, and the perpendicular and parallel components are computed separately. We also assume that the swimmer is either entirely an equipotential surface, or an entirely insulating surface such that only the first term in the Legendre-Fourier modes of the temperature expansion contributes to the speed generation, see~\cref{app:temp}. Though we should emphasize that this is not a limitation of the Teubner method. The laborious calculation is provided in~\cref{app:teubner}. Grouping the expressions for the individual contributions together, we obtain for the total speed of a swimmer
\begin{align}
\label{eq:speed_resultT} \bar{U} &= - \frac{k_{\mathrm{B}} T^{\infty} n^{\infty}}{6 \eta^{\infty} a} \left( \lambda^{\infty} \right)^{2} \left[ 8 \beta \phi(\vec{r}_{s}) - \left( 1 - \epsilon^{*} \right) \phi(\vec{r}_{s})^{2} \right] \bar{t}_{1}~, 
\end{align}
which matches the slip-layer result of~\cref{eq:speed_result}. 

\section{\label{sec:thick}The H{\"u}ckel Limit}

In this section, we use a regime-splitting approach to determine the leading-order departure from the salt-free limit. That is, we study systems in which the Debye length is large $\kappa^{\infty} a \ll 1$.

\subsection{\label{sub:problem}Splitting the Equation System}

We start from the equations for the equilibrium and non-equilbrium fields, of which the former are provided in~\cref{app:linearized} and the latter are given by Eqs.~\eqref{eq:Poissonlinneqred},~\eqref{eq:Xred}, and~\eqref{eq:dXred}. The equilibrium electric field obeys the regular Debye-H{\"u}ckel expression
\begin{align}
\label{eq:phieqlarge} \phi^{\mathrm{eq}}(\vec{r}) &= \phi_{s} \left( \frac{a}{r} \right) e^{- \kappa^{\infty} (r - a) }~,
\end{align}
where $\phi_{s}$ is a homogeneous surface potential, related either to the conducting or the insulating boundary condition. We introduce the splitting parameter $\xi = \kappa^{\infty} a \ll 1$ and the coordinate transform $\vec{r} \equiv a \vec{y}$ to approximate the equilibrium potential as
\begin{align}
\label{eq:phieqapproxpot} \phi^{\mathrm{eq}}(\vec{y}) &= \left\{ \begin{array}{cl} \tilde{\phi}_{s}/y & y < \xi^{-1} \\ 0 & y > \xi^{-1} \end{array} \right.~,
\end{align} 
up to order $\mathcal{O}(1)$. We have introduced the expansion prefactor $\tilde{\phi}_{s}$, which assumes the value $\tilde{\phi}_{s} = \phi_{0}$ for a conductor and $\tilde{\phi}_{s} = (a e \sigma_{0})/(k_{\mathrm{B}} T^{\infty} \epsilon^{\infty})$ for an insulator, respectively. Note that we assume that the potential is essentially unscreened inside the double layer and fully screened outside; we still require $x_{\pm}(\vec{y}) = \mp \phi^{\mathrm{eq}}(\vec{y})$. 

Using the above expressions and expanding Eqs.~\eqref{eq:Poissonlinneqred},~\eqref{eq:Xred}, and~\eqref{eq:dXred} to $\mathcal{O}(\xi)$, we obtain the following for $y < \xi^{-1}$:
\begin{align}
\label{eq:Poissonapprox} \nabla^{2} \phi^{\mathrm{neq}}_{\mathrm{in}}(\vec{y}) &= - \epsilon^{\ast} \vec{\nabla} t(\vec{y}) \cdot \vec{\nabla} \phi^{\mathrm{eq}}(\vec{y})~; \\
\label{eq:Xapprox} \nabla^{2} X^{\mathrm{neq}}_{\mathrm{in}}(\vec{y}) &= \beta \vec{\nabla} t(\vec{y}) \cdot \vec{\nabla} \phi^{\mathrm{eq}}(\vec{y})~; \\
\label{eq:dXapprox} \nabla^{2} \delta X^{\mathrm{neq}}_{\mathrm{in}}(\vec{y}) &= \left( 1 + \gamma + \epsilon^{\ast} \right) \vec{\nabla} t(\vec{y}) \cdot \vec{\nabla} \phi^{\mathrm{eq}}(\vec{y})~.
\end{align}
The relation $a \vec{\nabla}_{\vec{r}} = \vec{\nabla}_{\vec{y}}$ was used to obtain gradients and Laplacians in terms of $\vec{y}$; the subscript is dropped throughout for notational convenience. Note that the equations for all fields have the same shape, which can be solved.

\subsection{\label{sub:general}Solving the General Differential Form}

We write a general Legendre-Fourier series for the reduced temperature, mimicking the result obtained in~\cref{app:linearized}. That is,
\begin{align}
\label{eq:tempexp} t(\vec{y}) &= \sum_{j = 0}^{\infty} \check{t}_{j} y^{-(j+1)} P_{j} \left( \cos \theta \right)~,
\end{align}
with $P_{i}$ the $i$-th Legendre polynomial and the $\check{t}_{j}$ prefactors of the temperature expansion that can be related to the $\bar{t}_{j}$ provided in~\cref{app:temp}. Then, the general differential form associated with our splitting approach may be recast as
\begin{align}
\label{eq:general} \nabla^{2} G(\vec{y}) &= g \vec{\nabla} t(\vec{y}) \cdot \vec{\nabla} \phi^{\mathrm{eq}}(\vec{y})~,
\end{align}
with $G$ a function and $g$ some prefactor. The gradient of the equilibrium potential only has a radial component, under our constraining assumptions, therefore
\begin{align}
\label{eq:genexp} \nabla^{2} G(\vec{y}) &=  g \tilde{\phi}_{s} \sum_{j = 0}^{\infty} \check{t}_{j} (j + 1) y^{-(j+4)} P_{j} \left( \cos \theta \right)~.
\end{align}
A solution to this problem should also decompose into Legendre-Fourier modes, hence we make the \textit{ansatz}
\begin{align}
\label{eq:trial} G(\vec{y}) &= \sum_{j = 0}^{\infty} h_{j}(y) P_{j} \left( \cos \theta \right)~,
\end{align}
with $h_{j}$ functions to be determined. From Eqs.~\eqref{eq:genexp} and~\eqref{eq:trial} it then follows that the $h_{j}$ satisfy
\begin{align}
\label{eq:diffh} h_{j}''(y) + \frac{2}{y}h_{j}'(y) - \frac{j(j+1)}{y^{2}}h_{j}(y) &=  g \tilde{\phi}_{s} \check{t}_{j} \frac{(j + 1)}{y^{(j+4)}}~,
\end{align}
with the prime denoting the derivative with respect to $y$. These differential equations have solutions
\begin{align}
\label{eq:hj} h_{j}(y) &= \left( \frac{1}{2}  g \tilde{\phi}_{s} \check{t}_{j} + C_{j} y \right) y^{-(j+2)}~,
\end{align}
where the $C_{j}$ are constants of integration to be determined. We have removed the nonconvergent part in the limit of $y \uparrow \infty$, since we are interested in the limit $\xi^{-1} \uparrow \infty$ and the solutions should be bounded for all $\xi$.

\subsection{\label{sub:forcbal}The Thermo(di)electrophoretic Body Force}

Employing the general solution,~\cref{eq:hj}, we obtain solutions to Eqs.~\eqref{eq:Poissonapprox}~-~\eqref{eq:dXapprox} by imposing boundary conditions. We start with $\phi^{\mathrm{neq}}_{\mathrm{in}}$, which vanishes at the surface of the particle for an equipotential surface, leading to
\begin{align}
\nonumber \phi^{\mathrm{neq}}_{\mathrm{in}}(\vec{y}) &= \frac{1}{2} \sum_{j = 0}^{\infty} \check{t}_{j} \psi_{j}(y) P_{j} \left( \cos \theta \right)~; \\
\label{eq:phinonapppot} \psi_{j}(y) &= \epsilon^{\ast} \tilde{\phi}_{s} \left( y - 1 \right) y^{-(j+2)}~.
\end{align}
For an insulating surface, the derivative of $\phi^{\mathrm{neq}}_{\mathrm{in}}$ vanishes at the surface. This leads to the following non-equilibrium potential expansion coefficients
\begin{align}
\label{eq:phinonappcha} \psi_{j}(y) = - \epsilon^{\ast} \tilde{\phi}_{s} \frac{ (j + 1) - (j + 2) y }{(j+1)} y^{-(j+2)}~.
\end{align}

The fields $X^{\mathrm{neq}}_{\mathrm{in}}$ and $\delta X^{\mathrm{neq}}_{\mathrm{in}}(\vec{y})$ are solved for by matching the expansion to the outer boundaries at $\vert \vec{y} \vert = \xi^{-1}$, as expressed by $X^{\mathrm{neq}}_{\mathrm{in}}(\vec{y}) \vert_{y = \xi^{-1}} = -\gamma t(\vec{y}) \vert_{y = \xi^{-1}}$, and $\delta X^{\mathrm{neq}}_{\mathrm{in}}(\vec{y}) \vert_{y = \xi^{-1}} = -\beta t(\vec{y}) \vert_{y = \xi^{-1}}$. This results in
\begin{align}
\nonumber X^{\mathrm{neq}}_{\mathrm{in}}(\vec{y}) &= - \frac{1}{2}  \sum_{j = 0}^{\infty} \check{t}_{j} X_{j}(y) P_{j} \left( \cos \theta \right)~; \\
\label{eq:Xnonapp} X_{j}(y) &= \left( 2 \gamma y + \beta \tilde{\phi}_{s} \left( \xi y - 1 \right) \right) y^{-(j+2)}~,
\end{align}
and
\begin{align}
\nonumber \delta X^{\mathrm{neq}}_{\mathrm{in}}(\vec{y}) = - \frac{1}{2}  \sum_{j = 0}^{\infty} \check{t}_{j} \delta X_{j}(y) P_{j} \left( \cos \theta \right)~; \\
\label{eq:dXnonapp} \delta X_{j}(y) = \left( 2 \beta y + \left( 1 + \gamma + \epsilon^{\ast} \right) \tilde{\phi}_{s} \left( \xi y - 1 \right) \right) y^{-(j+2)}~,
\end{align}
for the non-equilibrium ionic strength and excess charge, respectively. Note that these are the same for insulators and conductors in the splitting formalism.

Outside the double layer ($y > \xi^{-1}$) we have $\phi^{\mathrm{neq}}_{\mathrm{out}}(\vec{y}) = -\beta t(\vec{y})$, $X^{\mathrm{neq}}_{\mathrm{in}}(\vec{y}) = -\gamma t(\vec{y})$, and $\delta X^{\mathrm{neq}}_{\mathrm{in}}(\vec{y}) = 0$, as before. This leads to a vanishing force acting on the fluid up to $\mathcal{O}(\tau^{3})$. Inside the screening layer, we write the force on the fluid in terms of $\vec{y}$ and $\xi$
\begin{align}
\nonumber \vec{f}_{\mathrm{in}}^{\mathrm{neq}}(\vec{y})&= - \frac{ \epsilon^{\infty} \left( k_{\mathrm{B}} T^{\infty} \right)^{2} }{ e^{2} a^{3} } \bigg[ \xi^{2} \delta X_{\mathrm{in}}^{\mathrm{neq}}(\vec{y}) \vec{\nabla} \phi_{\mathrm{in}}^{\mathrm{eq}}(\vec{y}) - \xi^{2} \phi_{\mathrm{in}}^{\mathrm{eq}}(\vec{y}) \vec{\nabla} \phi_{\mathrm{in}}^{\mathrm{neq}}(\vec{y}) \\
\label{eq:fstapprox} & \phantom{= \frac{ \epsilon^{\infty} \left( k_{\mathrm{B}} T^{\infty} \right)^{2} }{ e^{2} a^{3} } \bigg[} \quad + \frac{1}{2} \epsilon^{\ast} \left\vert \vec{\nabla} \phi_{\mathrm{in}}^{\mathrm{eq}}(\vec{x}) \right\vert^{2} \vec{\nabla} t(\vec{s}) \bigg]~.
\end{align}
Here, the temperature variation of the dielectric permittivity, as specified to first order by $\epsilon^{\ast}$, becomes the dominant term. This is expected, since thermo\textit{di}electrophoresis is the only effect contributing to the self-propulsion in the salt-free limit (without counterions) and results in a finite swim speed. 

\subsection{\label{sub:thickspeed}Swim Speed in the H{\"u}ckel Limit}

The solutions for the fields can be used in conjunction with Teubner's formalism, see~\cref{eq:Tspeed}, to obtain self-propulsion speeds in the H{\"u}ckel limit; admittedly, after laborious algebraic bookkeeping. The conducting swimmer has a speed
\begin{align}
\nonumber \tilde{U} &= - \frac{ \epsilon^{\infty} \epsilon^{\ast} \left( k_{\mathrm{B}} T^{\infty} \right)^{2} }{ 105 \eta^{\infty} e^{2} a } \phi_{0}^{2} \bar{t}_{1} \\
\label{eq:speed_huckel_pot} & \phantom{=} - n^{\infty} \frac{ a k_{\mathrm{B}} T^{\infty} }{ 180 \eta^{\infty} } \left( 30 \beta - \left( 4 + 4 \gamma + 15 \epsilon^{\ast} \right) \phi_{0} \right) \phi_{0} \bar{t}_{1}~,
\end{align}
while that of the insulating swimmer is given by
\begin{align}
\nonumber \tilde{U} &= - \frac{ \epsilon^{\infty} \epsilon^{\ast} \left( k_{\mathrm{B}} T^{\infty} \right)^{2} }{ 105 \eta^{\infty} e^{2} a } \tilde{\phi}_{s}^{2} \bar{t}_{1} \\
\label{eq:speed_huckel_cha} & \phantom{=} - n^{\infty} \frac{ a k_{\mathrm{B}} T^{\infty} }{ 360 \eta^{\infty} } \left( 60 \beta - \left( 8 + 8 \gamma + 45 \epsilon^{\ast} \right) \tilde{\phi}_{s} \right) \tilde{\phi}_{s} \bar{t}_{1}~,
\end{align}
with $\tilde{\phi}_{s} = (a e \sigma_{0})/(k_{\mathrm{B}} T^{\infty} \epsilon^{\infty})$. Note that here we have expanded the result in $\xi$, only retaining terms up to $\mathcal{O}(\xi^{3})$, and we have used $\check{t}_{1} = \bar{t}_{1}$. The identity holds for $j = 1$, but conversion factors are generally expected.

From the above equations, it is clear that even in the absence of salt, a polarization-based contribution to the swim speed remains. Burelbach and Stark similarly report a constant value of the speed for external thermoelectrophoresis in the H{\"u}ckel limit~\cite{burelbach2018determining}, which they refer to as a ``colloid hydration'' term. In addition, the direction of self-propulsion can change with ionic strength. Considering, for example the conducting swimmer, the leading term in the H{\"u}ckel limit $\tilde{U} \propto - \left( 30 \beta - \left( 4 + 4 \gamma \right) \phi_{0} \right) \phi_{0}$ may be smaller than zero ($\epsilon^{\ast} = 0$), whenever the leading term in the Smoluchowski limit $\tilde{U} \propto - \left( 8 \beta  - \phi_{0} \right) \phi_{0}$ is larger than zero and \textit{vice versa}, depending on the values of $\beta$ and $\gamma$. Lastly, whenever, $\epsilon^{\ast} = 0$, the limiting behavior at low salt concentration is that of a vanishing speed for both electrostatic boundary conditions. This is not the same as the dependency reported in Ref.~\citenum{burelbach2018determining}. The difference can be attributed to the geometry of the temperature field, which is not identical between self- and external thermoelectrophoresis in the H{\"u}ckel limit. Such geometric differences were recently showcased for electrophoresis~\cite{brown2017ionic}.

\section{\label{sec:intermediate}Intermediate Ionic Strengths}

Now that we have examined the swim speed in both limits of thin and thick screening layers, we can consider what happens in the intermediate regime. The algebra is rather complicated in this limit, hence we refer to our ESI$^\dag$ for the full details. The main idea is to not only assume a uniform electrostatic boundary condition, but also a specific form for the temperature profile. Only the first mode of the temperature expansion contributes to the speed. Thus, we restrict ourselves to the following temperature distribution $t(r,\theta) = -(a / r)^{2} \cos \theta$, where we introduce the minus sign to have a bottom-hot swimmer whenever $\Delta T > 0$. 

The relevant cross-coupling terms in Eqs.~\eqref{eq:Poissonlinneqred},~\eqref{eq:Xred}, and~\eqref{eq:dXred} may now be written as
\begin{align}
\label{eq:crosspois} t(r) \phi^{\mathrm{eq}}(\vec{r}) &= \tilde{C}_{0} \cos \theta \left( \frac{ a^{3} }{ r^{3} } \right) e^{-\kappa^{\infty} r} ; \\
\label{eq:crossflux} \vec{\nabla} t(\vec{r}) \cdot \vec{\nabla} \phi^{\mathrm{eq}}(\vec{r}) &= 2 \tilde{C}_{0} \cos \theta \left( \frac{ a^{3} }{ r^{5} } \right) \left( 1 + \kappa^{\infty} r \right) e^{-\kappa^{\infty} r}~,
\end{align}
where $\tilde{C}_{0}$ is a coefficient that accounts for all electrostatic prefactors---both for the equipotential and constant surface charge case---and we have made use of the uniformity of the imposed electrostatic boundary condition. The Poisson equation~\eqref{eq:Poissonlinneqred} reads
\begin{align}
\nonumber \nabla^{2} \phi^{\mathrm{neq}}(\vec{r}) + (\kappa^{\infty})^{2} \delta \! X^{\mathrm{neq}}(\vec{r}) &= \\
\label{eq:Poisson_plug} - \epsilon^{\ast} \tilde{C}_{0} \left( 2 + 2 \kappa^{\infty} r + \left( \kappa^{\infty} r \right)^{2} \right) \cos \theta \left( \frac{ a^{3} }{ r^{5} } \right) e^{-\kappa^{\infty} r}~, 
\end{align}
for the local salinity~\eqref{eq:Xred} we find
\begin{align}
\label{eq:X_plug} \nabla^{2} X^{\mathrm{neq}}(\vec{r}) &= 2 \beta \tilde{C}_{0} \left( 1 + \kappa^{\infty} r \right) \cos \theta \left( \frac{ a^{3} }{ r^{5} } \right) e^{-\kappa^{\infty} r}~,
\end{align}
and for the local charge excess~\eqref{eq:dXred} we obtain
\begin{align}
\nonumber \nabla^{2} \delta \! X^{\mathrm{neq}}(\vec{r}) + \nabla^{2} \phi^{\mathrm{neq}}(\vec{r}) = & \\
\label{eq:dX_plug} \tilde{C}_{0} \left( 2 (1 + \gamma) + 2 (1 + \gamma) \kappa^{\infty} r + \left( \kappa^{\infty} r \right)^{2} \right) \cos \theta \left( \frac{ a^{3} }{ r^{5} } \right) e^{-\kappa^{\infty} r}~. \quad &
\end{align}
Using \cref{eq:Poisson_plug}, we obtain a differential equation in terms of $\delta \! X^{\mathrm{neq}}(\vec{r})$ only, which is given by
\begin{align}
\nonumber \nabla^{2} \delta \! X^{\mathrm{neq}}(\vec{r}) - (\kappa^{\infty})^{2} \delta \! X^{\mathrm{neq}}(\vec{r}) &= \\
\nonumber \tilde{C}_{0} \left( 2 (1 + \gamma + \epsilon^{\ast} ) + 2 (1 + \gamma + \epsilon^{\ast}) \kappa^{\infty} r + (1 + \epsilon^{\ast}) \left( \kappa^{\infty} r \right)^{2} \right) & \\
\label{eq:dX_iso} \times \cos \theta \left( \frac{ a^{3} }{ r^{5} } \right) e^{-\kappa^{\infty} r}~. &
\end{align}
We have reduced the problem to an inhomogeneous Laplace equation for local non-equilibrium salinity $X^{\mathrm{neq}}(\vec{r})$, an inhomogeneous Helmholtz equation for the non-equilibrium charge excess $\delta \! X^{\mathrm{neq}}(\vec{r})$, and another inhomogeneous Laplace equation for the associated non-equilibrium potential $\phi^{\mathrm{neq}}(\vec{r})$; provided that a solution for $\delta \! X^{\mathrm{neq}}(\vec{r})$ has been established. This system of equations may be solved analytically, although no closed-form expressions can be obtained.

Taking the Smoluchowski limit, we arrive at $\lim_{\lambda^{\infty} \downarrow 0} \bar{U} = 0$ for the insulating surface and 
\begin{align}
\label{eq:speed_lim} \bar{U} &= - \frac{\left( k_{\mathrm{B}} T^{\infty}\right)^{2} \epsilon^{\infty}}{12 e^{2} \eta^{\infty} a} \left[ 8 \beta \phi_{0} - \left( 1 - \epsilon^{*} \right) \phi_{0}^{2} \right]~, 
\end{align}
both of which are in agreement with~\cref{eq:speed_result}. Unfortunately, using the full analytic approach makes it difficult to establish how the speed departs from these limits, see the ESI$^\dag$. The reason is that expressions appear in the solution that are problematic to evaluate numerically, as they invovle the near cancellation of large terms to give rise to a small, yet relevant values.

In the opposite (H{\"u}ckel) limit, we obtain for a conducting swimmer the following speed
\begin{align}
\nonumber \tilde{U} &= - \frac{ \epsilon^{\infty} \epsilon^{\ast} \left( k_{\mathrm{B}} T^{\infty} \right)^{2} }{ 105 \eta^{\infty} e^{2} a } \phi_{0}^{2} - \frac{ 2 \epsilon^{\infty} \epsilon^{\ast} \left( k_{\mathrm{B}} T^{\infty} \right)^{2} }{ 105 \eta^{\infty} e^{2} } \kappa^{\infty} \phi_{0}^{2} \\
\label{eq:speed_huckel_new_pot} & \phantom{=} - n^{\infty} \frac{ a k_{\mathrm{B}} T^{\infty} }{ 2520 \eta^{\infty} } \left( 420 \beta - \left( 49 - 161 \gamma - 16 \epsilon^{\ast} \right) \phi_{0} \right) \phi_{0} ~.
\end{align}
This expression has a similar shape as~\cref{eq:speed_huckel_pot}, barring a $\kappa^{\infty} \propto \sqrt{n^{\infty}}$ dependent term. This term comes from the outer region of the solution, where we had assumed a fully screened potential and set the force density to zero in~\cref{sec:thick}. We conclude that splitting gives an impression of the limit and some aspects of the departure therefrom, but can lead to qualitatively incorrect scaling. Nonetheless, when $\epsilon^{\ast} = 0$, Eqs.~\eqref{eq:speed_huckel_pot}~and~\eqref{eq:speed_huckel_new_pot} agree semi-quantitatively, with only minor changes in the prefactors. Therefore, the method can have merit whenever there is no contribution to self-propulsion outside the screening layer.

The insulating swimmer's speed in the H{\"u}ckel limit is given by
\begin{align}
\nonumber \tilde{U} &= - \frac{ a \epsilon^{\ast} }{ 105 \eta^{\infty} \epsilon^{\infty} } \sigma_{0}^{2} - \frac{ a^{3} e^{2} }{ 360 k_{\mathrm{B}} T^{\infty} \eta^{\infty} \left( \epsilon^{\infty} \right)^{2} } \left( 7 - 23 \gamma  \right) n^{\infty} \sigma_{0}^{2} \\
\label{eq:speed_huckel_new_cha} & \phantom{=} - \frac{ a^{2} e }{ 6 \eta^{\infty} \epsilon^{\infty} } n^{\infty} \sigma_{0}^{2} + \frac{ 8 a^{3} e^{2} \epsilon^{\ast} }{ 315 k_{\mathrm{B}} T^{\infty} \eta^{\infty} \left( \epsilon^{\infty} \right)^{2} } n^{\infty} \sigma_{0}^{2}~.
\end{align}
This expression also differs from the one provided in~\cref{sub:thickspeed} for the same reasons. We will show next that the speeds reported in this section do capture many of the speed features obtained using FEM, despite the limitations of the method.

\section{\label{sec:result}Numerical Results}

In this section, we discuss our numerical FEM results and show that these correspond to the expressions of our analytic calculations in the appropriate limits. We will predominantly use dimensionful units to make the connection with experiments and to highlight those regimes wherein we expect measurable results. 

\subsection{\label{sub:parameters}Parameter Choices}

Throughout, we assume a colloidal particle diameter of $\SI{1}{\micro\meter}$. We consider three types of swimmer material for the hot swimmers: no thermal conductivity contrast with water $K = 1$; polystyrene (PS), $K_{in} \equiv k_{\mathrm{PS}} / k_{f} = 0.0847$; and silica (\ce{SiO2}) $K_{\ce{SiO2}} = 2.34$. For the fluid, we use the physical properties of water at $T^{\infty} = \SI{298.15}{\kelvin}$ (room temperature): $\epsilon^{\infty} = \epsilon_{0} \epsilon_{r}$ with $\epsilon_{0}$ the vacuum and $\epsilon_{r} = 78.4$ the relative permittivity, $\eta^{\infty} = \SI{8.9e-4}{\pascal\second}$, and $k^{\infty} = k_{f} = \SI{0.591}{\watt\per\meter\per\kelvin}$~\cite{haynes2009handbook}. The ambient pressure is specified to be $p^{\infty} = \SI{1e5}{\pascal}$, approximately one atmosphere.

We further consider three types of ions to determine the effect of thermoelectrophoresis, one cation, sodium \ce{Na+}, and two anions, chloride \ce{Cl-} and hydroxide \ce{OH-}. This choice is based on the commonplaceness of these ions, as well as the fact that the \ce{Cl-} anion has a much smaller Soret coefficient than \ce{OH-}, allowing us to probe the effect thereof on the motion of the swimmer. The ionic diffusion coefficients are $D_{\ce{Na+}} = \SI{1.3e-9}{\meter\squared\per\second}$, $D_{\ce{Cl-}} = \SI{2.0e-9}{\meter\squared\per\second}$, and $D_{\ce{OH-}}^{\infty} = \SI{5.3e-9}{\meter\squared\per\second}$~\cite{samson2003calculation}. The thermal diffusion coefficients are given by $\alpha_{\ce{Na+}}^{\infty} = 0.7$, $\alpha_{\ce{Cl-}}^{\infty} = 0.1$, $\alpha_{\ce{OH-}}^{\infty} = 3.4$~\cite{eastman1928theory, agar1989single, majee2012charging}. In all cases, we set $\epsilon^{\ast} = 0$, dropping any thermal polarization effects, in order to facilitate the discussion of the results; the actual value $\epsilon^{\ast} \approx -1.3$, see~\cref{app:justify}.

Lastly, in the following we will consider both fully insulating and fully conducting surfaces. This abstraction best showcases the differences between these two electrostatic boundary conditions. In reality a gold- or carbon-coated $\mathrm{PS}$/\ce{SiO2} swimmer will (presumably) have some combination of these boundary conditions,~\textit{e.g.}, also see Ref.~\citenum{ly2018nanoscale}. However, for the purpose of quantifying the difference between these two cases, we will only consider the pure boundary conditions and several relevant values of the surface potential and charge, respectively. This facilitates direct comparison to our analytic results. We do not concern ourselves with the experimental realizability of these `pure' boundary conditions here.

\subsection{\label{sub:temperature}The Temperature Profile and Thermocharge}

\begin{figure}[!htb]
  \centering
  \includegraphics[width=3.3in]{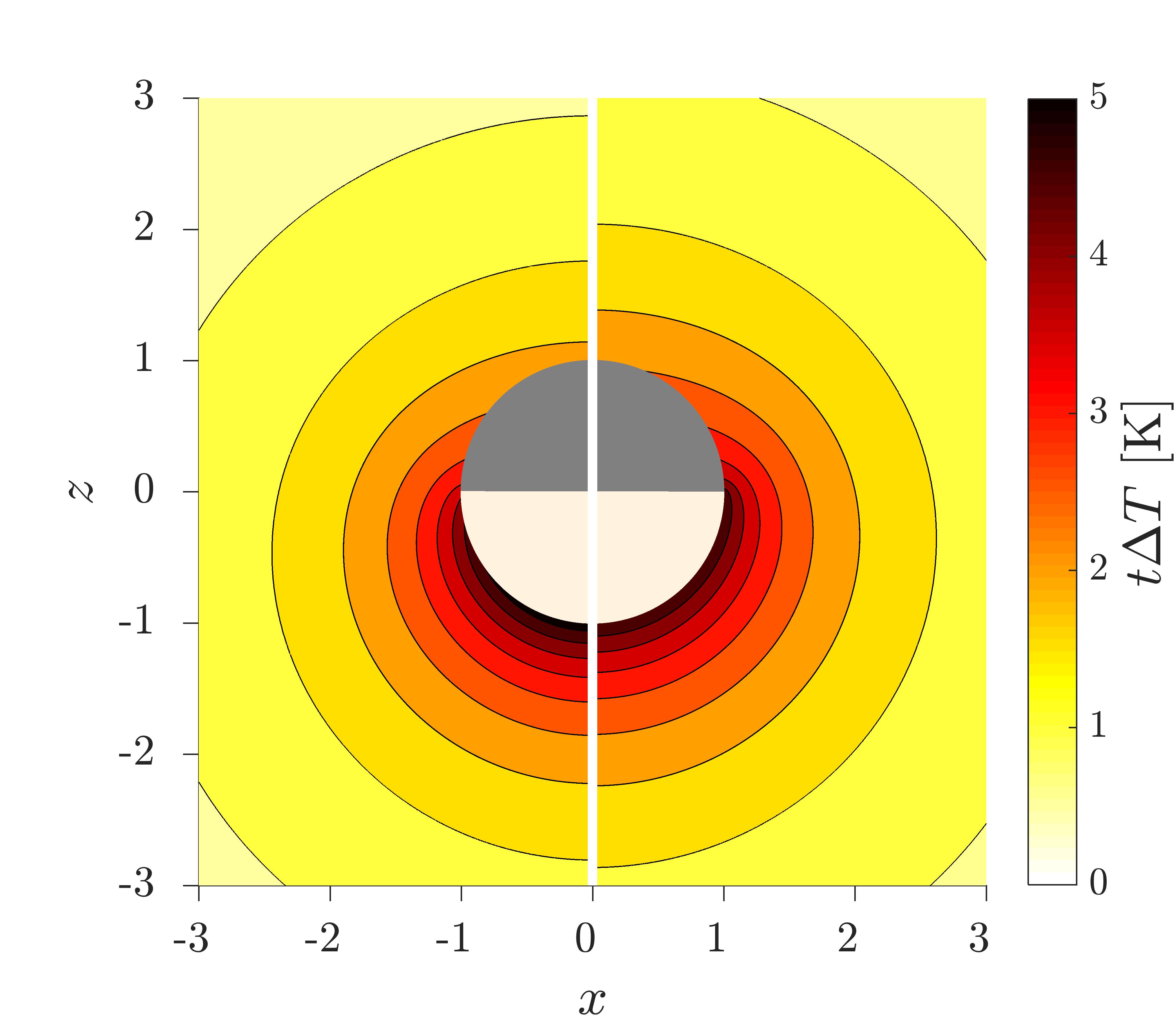}
  \caption{\label{fig2}Contour plot of the excess temperature $t \Delta T$ around a hot ($K = 1$) Janus swimmer in the $xz$-plane. In the left halfplane, we show $t$ when the heat flux $Q$ is fixed on the cap (white, $z \le 0$), such that the maximum temperature on the bottom hemisphere is $\approx \SI{5}{\kelvin}$. The right halfplane shows the temperature field with $\Delta T = \SI{5}{\kelvin}$ fixed on the heated cap.}
\end{figure}

Let us first examine the temperature profile around a heated ($K = 1$) swimmer. \cref{fig2} shows the temperature excess for both types of thermal boundary condition, where we chose the heat flux $Q$ such that the maximum deviation from the reservoir temperature, $\Delta T \approx \SI{5}{\kelvin}$, is comparable to the imposed excess temperature for the equi-temperature surface, $\Delta T = \SI{5}{\kelvin}$. The two temperature fields differ only slightly, and we will therefore focus on heat-flux boundary conditions in the following, unless stated otherwise. 

\begin{figure*}[!htb]
  \centering
  \includegraphics[width=4.95in]{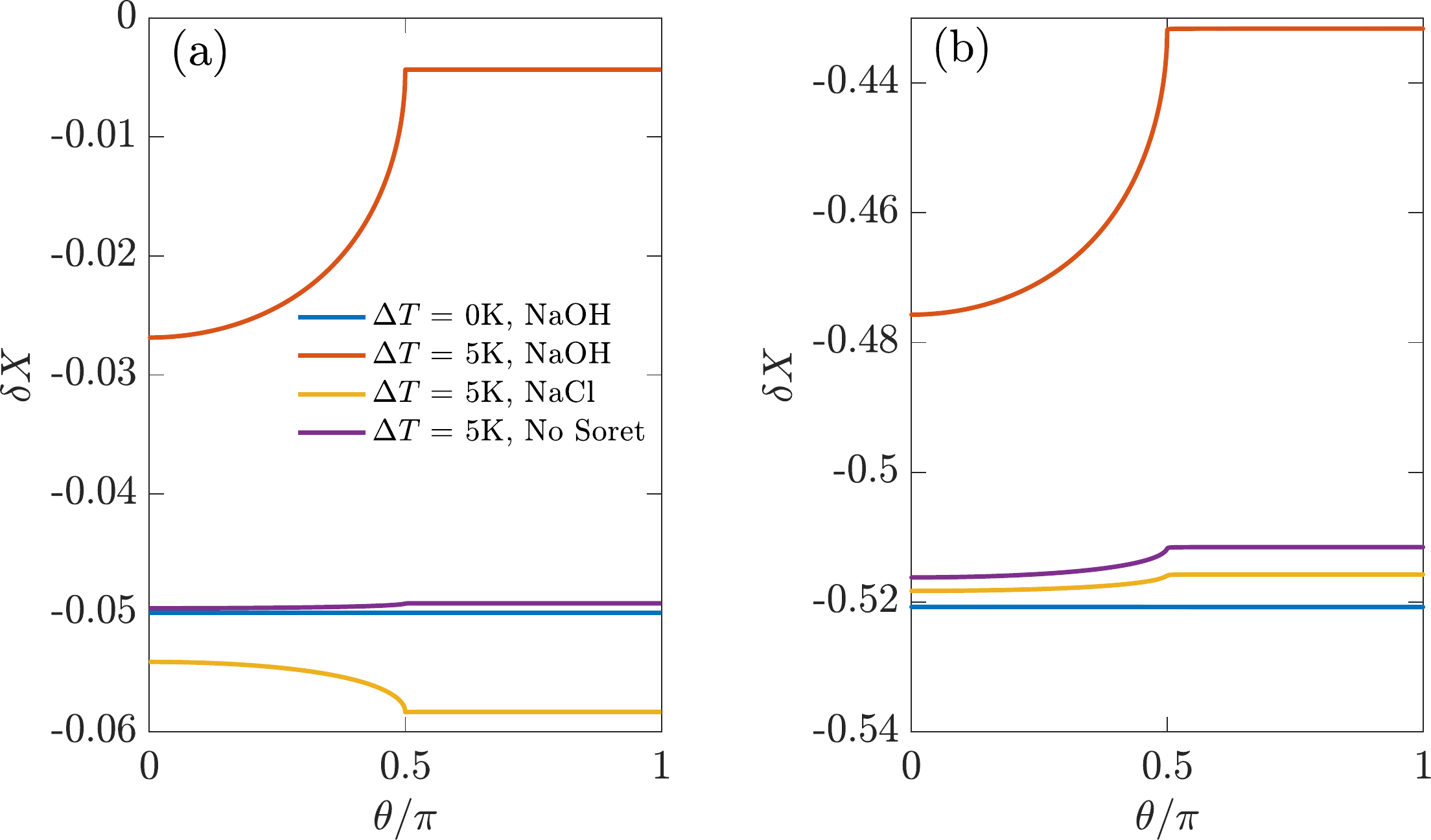}
  \caption{\label{fig3}The net charge $\delta \! X = \delta \! X^{\mathrm{eq}} + \tau \delta \! X^{\mathrm{neq}}$ along the particle contour parameterized by the polar angle $\theta$. The curves are for a Janus swimmer with $\Delta T \approx \SI{5}{\kelvin}$, $K = 1$, and an equipotential surface with $\phi_{0} = 0.05$ (a) and $\phi_{0} = 0.5$ (b); this corresponds to $\approx \SI{2.6}{\milli\volt}$ and $\approx \SI{13}{\milli\volt}$, respectively.}
\end{figure*}

Next, we turn our attention to the net charge at the surface of a hot ($K = 1$) swimmer with equipotential boundary condition, see~\cref{fig3}. When the particle is not heated ($\Delta T = \SI{0}{\kelvin}$), $\delta \! X$ is fixed, and equal and opposite to the imposed value of $\phi_{0}$ in our approximation. The agreement is good for $\phi_{0} = 0.05$ (in the linear regime), but there is an appreciable nonlinear effect for $\phi = 0.5$. The above nonlinearity can be better captured analytically by using Poisson-Boltzmann theory~\cite{majee2012charging, ly2018nanoscale}. However, most of the analytic manipulation performed in this paper cannot be accomplished in this more general case; the expressions become unwieldy.

Heating of the particle in a $\SI{1}{\milli\mole\per\liter}$ \ce{NaOH} solution leads to an increase in the anion concentration at the hot surface. Recall that for $\epsilon^{\ast} = 0$ the thermocharge at the surface is given by $\delta \! X^{\mathrm{neq}} = (\phi_{0} - \beta) t$ to first order, see~\cref{eq:dXinEPsol}. Here, $\beta = -2.7$ and $\phi_{0} = 0.05$, which gives $\delta \! X^{\mathrm{neq}} = 2.75 t$, and $\phi_{0} = 0.5$, which gives $\delta \! X^{\mathrm{neq}} = 3.2 t$, respectively. Hence, we expect $\delta \! X$ to increase at the heated cap --- it is nearly constant over that hemisphere --- and to be minimal at the pole of the particle, where the surface temperature is the lowest. Our linearized theory is qualitatively correct for both potentials, and we have quantitative agreement for $\phi = 0.05$.

The thermocharging effect is much smaller for a $\SI{1}{\milli\mole\per\liter}$ \ce{NaCl} solution due to the smaller Soret coefficient of the \ce{Cl-} anion ($\beta = 0.6)$; here we find $\delta \! X^{\mathrm{neq}} = (\phi_{0} - \beta)t = ( 0.05 - 0.6 ) t = -0.55 t$ and $( 0.5 - 0.6 ) t = -0.1 t$, respectively. In the linear regime, our theory predicts the correct sign change of the thermocharge with respect to the \ce{NaOH} solution, but in the nonlinear regime there is no qualitative agreement. The reason for this is revealed by examining the situation where \textit{no} Soret effect is included (purple curve). Here, we should obtain $\delta \! X^{\mathrm{neq}} = \phi_{0} t = 0.05 t$ and $0.5 t$, respectively. Clearly, the effect of nonlinearity is much stronger for the $\beta = 0$ thermocharging. 

\subsection{\label{sub:flow}The Flow Field around the Hot Swimmer}

One of the most important properties of the swimmer is the flow field generated by the non-equilibrium effect, as this governs to first order the interaction of the swimmer with its environment. This aspect was previously explored by Bickel~\textit{et al.}~\cite{bickel2013flow} for a hot swimmer that had a Seebeck-related slip velocity. Here, we include all terms leading to thermoelectric fluid motion in our equations and go beyond the Smoluchowski limit using FEM. 

\begin{figure}[!htb]
  \centering
  \includegraphics[width=0.9\columnwidth]{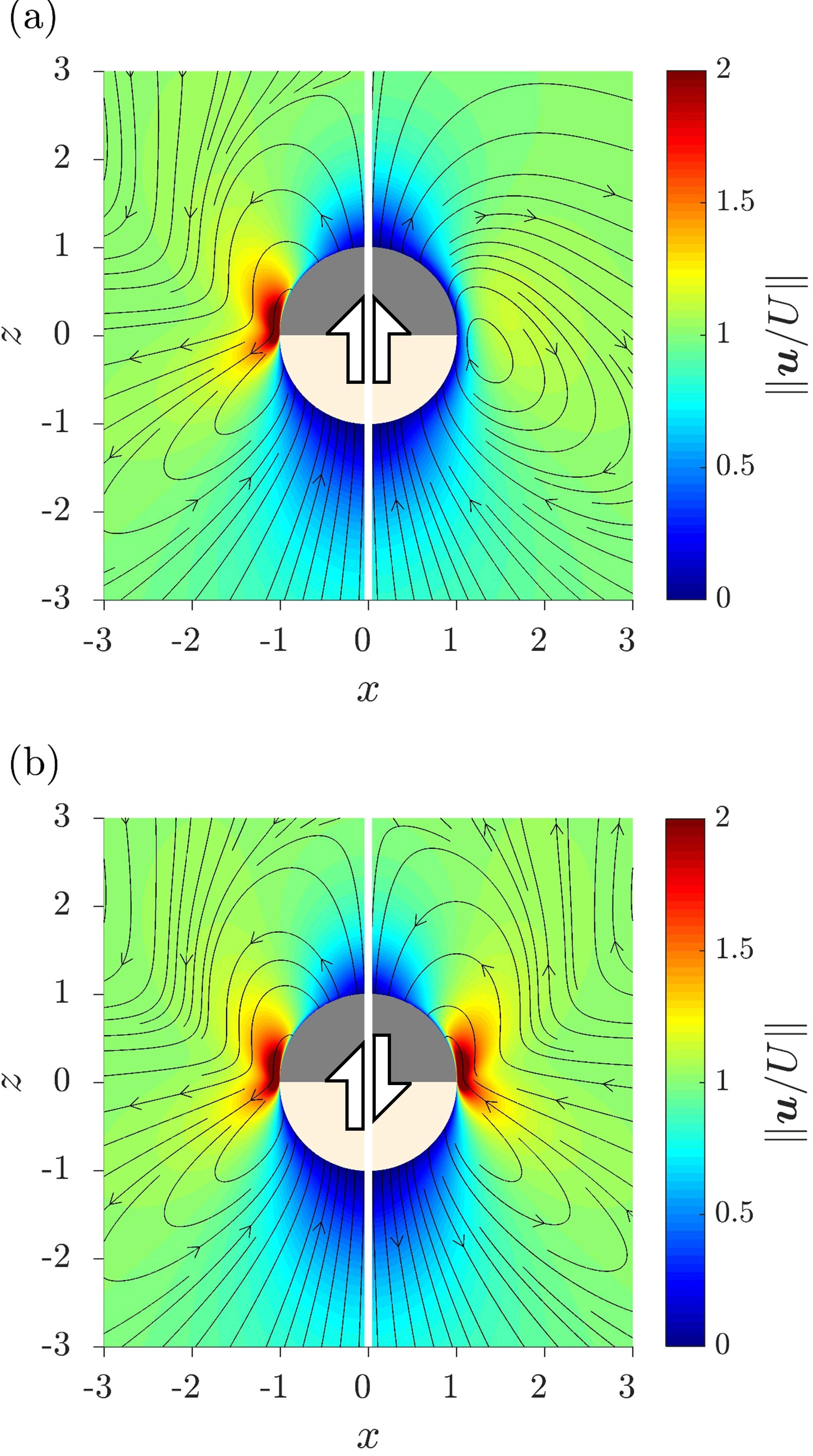}
  \caption{\label{fig4}Fluid velocity magnitude $\lVert \vec{u} \rVert$ divided by the absolute swim speed $\vert U \vert$ and streamlines for several hot ($K = 1$)  swimmers in the laboratory frame of reference. In (a), the electrostatic potential is fixed at $\phi_{0} = 0.5$ ($\approx \SI{13}{\milli\volt}$) and we impose $\Delta T = \SI{5}{\kelvin}$ at the heated cap. The concentration of \ce{NaOH} is $\SI{1}{\milli\mole\per\liter}$ is the left panel and $\SI{1e-03}{\milli\mole\per\liter}$ in the right panel. In (b), the surface charge of the particle is fixed at $\SI{5e-3}{\elementarycharge\per\nano\meter\squared}$ and $\Delta T = \SI{5}{\kelvin}$. The electrolyte in the left panel is $\SI{1}{\milli\mole\per\liter}$ \ce{NaOH} and, while in the right panel it is $\SI{1}{\milli\mole\per\liter}$ \ce{NaCl}. Notice the opposite direction of the streamlines in the two panels as the swimmers translate in opposite directions. The large arrows in the center of the swimmer indicate the direction of motion.}
\end{figure}

\cref{fig4} shows representative flow fields for several swimmer and environmental configurations. We find that by lowering the salinity the puller type flow is suppressed, leaving a more neutral-squirmer flow field, see~\cref{fig4}a for the effect for a hot swimmer in a \ce{NaOH} solution. Changing the anion type and leaving the other parameters the same can be used to change the direction of motion and to change from a puller- to a pusher-type flow field, thereby strongly modifying the interaction of the hot swimmer with its environment.

\subsection{\label{sub:thermcond}Thermal Conductivity and Soret Coefficients}

We start by providing the the dimensionful expressions for the thermoelectrophoretic self-propulsion speed in the thin-screening-layer limit here. These can be obtained by multiplying $\bar{U}$ (\cref{eq:speed_result}) with $\Delta T/T^{\infty}$ and utilizing the expressions from~\cref{app:temp}:
\begin{align}
\nonumber U &= - \frac{k_{\mathrm{B}} T^{\infty} n^{\infty}}{6 \pi \eta^{\infty} a} \left( \lambda^{\infty} \right)^{2} \left[ 8 \beta \phi_{i} - \phi_{i}^{2} \right] \\
\label{eq:SPEED} &\phantom{=} \times \left\{ \begin{array}{cl} \displaystyle \frac{\Delta T}{ T^{\infty} } & \mathrm{isothermal~cap} \\ & \\ \displaystyle \frac{3 \pi a Q}{4 k_{\mathrm{f}} (2 + K) T^{\infty}} & \mathrm{constant~heat~flux~cap} \end{array} \right. ~. 
\end{align}
Note that the dimensionful expression for the constant heat flux condition $Q$ is not dependent on the particle radius $a$ here. However, this is not the case in practice, since typically $q$ dependents on $a$. In general, $q \propto I \sigma_{\mathrm{abs}}/a^2$, where $I$ is the illumination intensity and $\sigma_{\mathrm{abs}}$ is the absorption cross section. The dependence of $\sigma_{\mathrm{abs}}$ on $a$, however, is non-trivial. $\sigma_{\mathrm{abs}}\propto a^3$ for small particles with $a\sim \mathcal{O}\left(\SI{0.01}{\micro\meter}\right)$, while for big particles with $a\sim \mathcal{O}\left(\SI{10}{\micro\meter}\right)$, $\sigma_{\mathrm{abs}}\propto a^2$ \cite{bregulla2015size}. Therefore, $q$ varies from $q\propto I/a$ to $q\propto I$, while being more complex in between. We only consider a fixed value of $a$ here and will ignore such dependencies in the following.

\begin{figure}
  \centering
  \includegraphics[width=\columnwidth]{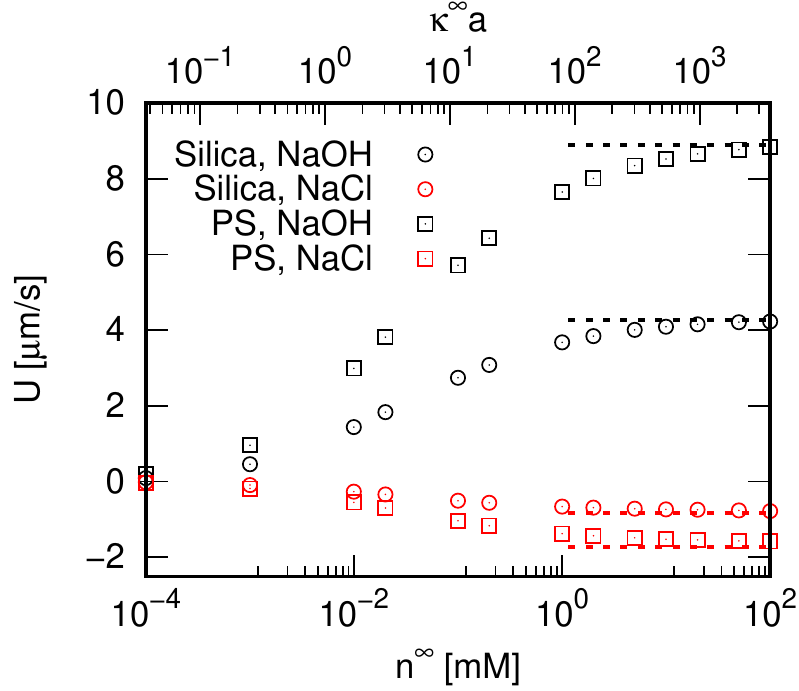}
  \caption{\label{fig5}Swimmer speed $U$ as a function of the bulk salt concentration $n^{\infty}$ for two electrolytes, \ce{NaOH} (black) and \ce{NaCl} (red), and two materials which comprise the hot swimmer, \ce{SiO2} (circles) and \ce{PS} (squares). As before, $\phi_{0} = 0.5$ ($\approx \SI{13}{\milli\volt}$) and we used a constant heat-flux boundary condition such that $\Delta T = \SI{5}{\kelvin}$. The Debye length decreases towards the right and dashed lines indicate the analytic limit $\lambda^{\infty} \downarrow 0$.}
\end{figure}

\Cref{fig5} shows the swim speed as a function of the bulk salt concentration for four representative swimmer/salt combinations and an equipotential boundary condition. The effect of the difference in thermal conductivity is quantitative, leading to an appreciable increase in absolute speed with reduced $K$. The direction of swimming is reversed between the two types of salt, as shown in~\cref{fig4} and previously reported by Ly~\textit{et al.}~\cite{ly2018nanoscale}. In all cases we obtain significant swimming speeds, $\mathcal{O}\left(\SI{1}{\micro\meter\per\second}\right)$, in physiological to high salt concentrations. 

Note that we accurately capture the analytic limit for our equipotential swimmer, even though we do not resolve the thermocharge correctly, see~\cref{fig3}. In the analytic theory we find that for such a swimmer $U \propto n^{\infty} \left( \lambda^{\infty} \right)^{2} \propto 1$ (in terms of $n^{\infty}$). Our result implies that the swim speed is independent of the reservoir concentration to first order. This is borne out by our numerical data in~\cref{fig5}, which is almost constant over a large range in $n^{\infty}$. Higher-order terms would capture the departures from the constant value of $U$ close to the limit $\lambda^{\infty} \downarrow 0$. However, it is non-trivial to analyze these, as follows from~\cref{sec:thick,sec:intermediate}.

The physical interpretation of the near-constant value of the speed is that smaller fluid velocities can be generated in a thinner screening layer. However, this is exactly counterbalanced by the increased steepness of the electrostatic potential therein, which in itself leads to higher speeds. Whenever $\beta = 0$, the ion variation is in the bulk couples back to the surface, resulting in a dependency $U \propto \phi_{0}^{2}$, but with the same constancy in $n^{\infty}$.

\subsection{\label{sub:potvssurf}Conducting and Insulating Hot Swimmers}

\begin{figure}[!htb]
  \centering
  \includegraphics[width=1.0\columnwidth]{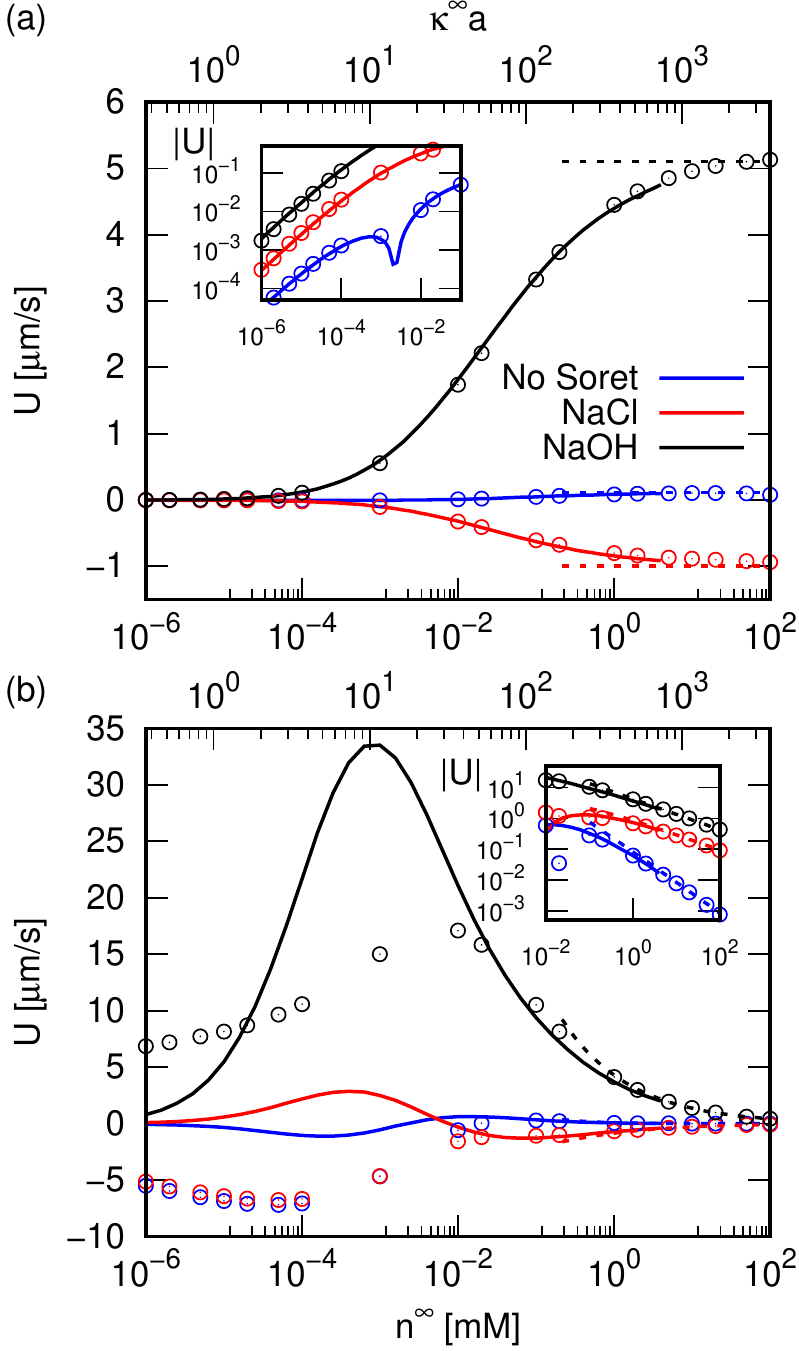}
  \caption{\label{fig6}Silica swimmer speed $U$ as a function of salt concentration $n^{\infty}$ for two electrolytes, \ce{NaCl} (red) and \ce{NaOH} (black), and when the Soret effect is neglected ($\beta = \gamma = 0$; blue). Symbols indicate the FEM result, solid curves show the analytic theory of~\cref{sec:intermediate}, dashed lines the departure from the limiting value (\cref{eq:speed_result}). In (a) we use an equipotential electrostatic boundary condition with $\phi_{0} = 0.5$ ($\approx \SI{13}{\milli\volt}$) while in (b) we used a constant surface charge one $\sigma = \SI{5e-3}{\elementarycharge\per\nano\meter\squared}$. In both panels $\Delta T = \SI{5}{\kelvin}$ is fixed at the heated cap. The dashed lines indicate the prediction of~\cref{eq:speed_result}. The insets show the agreement between theory and numerical results in the H{\"u}ckel limit (a) and the Smoluchowski limit (b).}
\end{figure}

\cref{fig6} shows the effect of the electrostatic boundary condition and the Soret effect on the motion of the hot swimmers as a function of the bulk salt concentration. Comparing the two panels of~\cref{fig6} the impact of the surface properties on the swim speed becomes evident. Equipotential swimmers have nearly constant swim speed in the thin-screening-layer limit, see~\cref{fig6}a, which we commented on in~\cref{sub:thermcond}. However, the speed of an insulating swimmer drops to zero in this limit. Note the excellent agreement between our analytic expressions and the FEM results here, see the inset to~\cref{fig6}b.

The speed of an insulating swimmer in the Smoluchowski regime is physically interpreted as follows. The surface potential $\phi_{s}$ corresponding to this boundary condition varies with the ion concentration,~\textit{i.e.}, $\phi_{s} \propto \lambda^{\infty}$. Thus, $U \propto n^{\infty} \left( \lambda^{\infty} \right)^{3} \propto \left( n^{\infty} \right)^{-1/2}$ ($\beta \ne 0)$ and $U \propto \left( n^{\infty} \right)^{-1}$ ($\beta = 0$) to first order, as can be appreciated from~\cref{fig6}b. The intuition is that, the surface potential must decrease with the Debye length in order to maintain the gradient-based boundary condition. Consequently, the coupling between the electric field and the temperature-induced ion currents reduces proportionally, leading to a vanishing speed.

Turning to the opposite limit of $n^{\infty} \downarrow 0$, an equipotential swimmers' speed drops to zero, see~\cref{fig6}a. This agrees with the result of~\cref{eq:speed_huckel_pot} which predicts a dependence $U \propto n_{0}$. In fact, the inset to \cref{fig6}a shows that the agreement is even quantitative. The physical interpretation of this scaling is as follows. Any $n^{\infty}$ perturbation the unscreened potential, will predominantly generate an out-of-equilibrium ion profile, rather than directly screen the potential, which results in a linear dependence.

Interestingly, insulating swimmers have sizable speeds for low ionic strengths, see~\cref{fig6}b, which are only weakly dependent on the salt concentration over several decades. From our analytic theory it would follow that the speed decays to zero in the limit $n^{\infty} \downarrow 0$, assuming that the nonlinearities in the electrostatic potential do not play a significant role. We therefore suspect that \cref{fig6}b shows only a part of the decay. Clearly, the nonlinearities do play a role, as the prediction of our analytic theory is not qualitative in for low salt concentrations.

Establishing the limiting value numerically proved problematic, due to the computationally demanding nature of such an FEM calculation. The same holds for solving the differential equation system using other numerical solvers, though Burelbach and Stark have made progress for external thermodielectrophoresis~\cite{burelbach2018determining}. However, we would like to note that in a real system the no-salt limit cannot be achieved, due to water autoionization, as well as \ce{CO2} (typically) dissolving in the medium. This makes determining the limit somewhat academic, especially as we have ignored $\epsilon^{\ast}$ contributions.

Finally, observe that mobility reversals that are present both in our FEM calculations and our analytic theory for both conducting and insulating swimmers. This reversal is best observed for the former in the inset to~\cref{fig6}a. Such reversals are reminiscent of external electrophoresis~\cite{obrien1978electrophoretic} and presumably have the same non-linear origin. This is why they only show up for $\beta = 0$ in the case of a conducting swimmer. For external thermoelectrophoresis, Burelbach and Stark observe a similar inversion as a function of the Debye length~\cite{burelbach2018determining}, which could be attributed to the same mechanism. However, we wish to emphasize that there are geometric differences between self- and external thermoelectrophoresis, hindering a direct comparison of the mobility inversion.

\section{\label{sec:conclusion}Discussion and Outlook}

Summarizing, we numerically determined the self-thermo(di)electrophoretic propulsion speed of a hot swimmer for various boundary conditions and environmental parameters. Specifically, we examined the largely unexplored regime of wide electrostatic screening layers (low ionic strength) using the finite-element method and verified our results in the appropriate limits using linear analytic theory. We discussed in depth the limitations of the Debye-H{\"u}ckel approximation for the case of an insulating swimmer in the electrostatic H{\"u}ckel limit. The strengths and weaknesses of various approaches have hereby been thoroughly charted and suggest that more involved non-linear calculations are required in some cases. Fortunately, in spite of relatively strong reductions, we were able to (semi-)quantitatively capture the speed dependence found by FEM in the H{\"u}ckel limit for conducting swimmers. In fact, our expressions hold up to reasonably high values of the surface potential.

Turning to the physics, we obtained $\si{\micro\meter\per\second}$ swimming speeds in physiological salt concentrations $n^{\infty} \gtrsim \SI{1}{\milli\mole\per\liter}$ for an equipotential boundary condition. These speeds are nearly independent of the salt concentration, in the thin screening-layer limit, due to a cancellation of ion-dependencies. For an insulating swimmer, however, propulsion speeds are low in this regime and they drop off with increasing bulk salinity. Counterintuitively, the speeds for an insulating swimmer appear to be nearly constant and are considerable in the limit of large Debye lengths, even without the thermo\textit{di}electrophoretic effect taken into account. Between these two limits the direction of self-propulsion can change, as clearly evidenced by our FEM result. This finding is supported by our analytic results, but the most striking inversion is not well captured.

To the best of our knowledge the low-salinity limit has not yet been systematically explored experimentally. Here, we have shown here that there are potentially high swimming speed and interesting nonlinearities to be found in this limit. In the real world, gold- or carbon-coated hot swimmers can possess more complex electrostatic boundary conditions than we have considered. These hot Janus swimmers may be partially conducting and partially insulating or have some intermediate form~\cite{ly2018nanoscale}. It is conceivable that this leads to a ``best of both worlds'' scenario, where a relatively high swim speed is maintained over all values of $n^{\infty}$. How such non-uniformity impacts the above results in the thick-screening-layer limit will be left to future study.

\textit{Acknowledgements} --- We thank Marie Sk{\l}odowska-Curie Intra European Fellowship (G.A.~No.~654916, JdG, and~656327, SS) within Horizon 2020 for funding. JdG further acknowledges funding through association with the EU-FET project NANOPHLOW (766972) within Horizon 2020. We are grateful to Aidan Brown, Mathijs Janssen, and Ben Werkhoven for fruitful discussions.

\scriptsize{

\providecommand*{\mcitethebibliography}{\thebibliography}
\csname @ifundefined\endcsname{endmcitethebibliography}
{\let\endmcitethebibliography\endthebibliography}{}

} 

\appendix

\section{\label{app:justify}Justification of Linearization}

In this appendix, we justify the reductions we made in the main text. We refer to the work by Dietzel and Hardt~\cite{dietzel2017flow} and references therein for a full discussion of the first-order Taylor expansion coefficients to the physical quantities. Here, we reproduce the values listed in Ref.~\citenum{dietzel2017flow} in terms of our notation. For the medium they found $\eta^{\ast} \approx -5$, $k^{\ast} \approx 0.7$, and $\epsilon^{\ast} \approx -1.3$. For the ``typical'' ions \ce{Na+}, \ce{K+}, and \ce{Cl-} they obtained $D^{\ast}_{\pm} \approx 6$~\cite{dietzel2017flow}. For the variation in the thermal diffusion coefficient of the ions only limited data is available in the literature. We refer to the work of Caldwell~\cite{caldwell1973thermal}, from which we obtain $D^{\ast}_{\pm} \approx 1$ for \ce{NaCl} and a temperature dependence for the thermal diffusivity given by $\alpha_{\pm}^{\ast} \approx 2$. There is clearly some variation in the literature values, but importantly all these numbers are order unity and we are therefore justified in ignoring these temperature dependencies. They come into the differential equations for the potential and concentration at $\mathcal{O}(\tau^{2})$, with $\tau^{2} \approx 3\cdot10^{-4}$ for $\Delta T \le \SI{5}{\kelvin}$, leading to minute variations. For the speed they come into the expression at $\mathcal{O}(\tau)$, which leads to a change of at most $10\%$. The only exception to this rule is $\epsilon^{\ast}$, which enters the theory at linear order and contributes as a constant to the speed, see~\cref{eq:SPEED}. Finally, the P{\'e}clet number for the ions in our system is given by $\mathrm{Pe} = U a/D$, with $U \le \SI{10}{\micro\meter\per\second}$ the typical velocity, $a = \SI{1}{\micro\meter}$ the radius of the colloid, and $D \ge \SI{1.0e-9}{\meter\squared\per\second}$ the smallest ion diffusion coefficient for convenience. Using the numbers provided in~\cref{sec:result}, we find that $\mathrm{Pe} \le 10^{-2}$, therefore we can safely ignore advective terms in~\cref{eq:conslinneq}. Similarly, we can ignore thermal advection terms, since thermal diffusivities are orders of magnitude larger than typical ion diffusion coefficients.

\section{\label{app:temp}Temperature Profiles}

We follow Ref.~\citenum{bickel2013flow} to obtain the temperature profiles in our system and reproduce their results here in our notation for completeness. We examine two boundary conditions for the coated hemisphere. For the a cap maintained at constant temperature, we assume for the thermal conductivities $k_{\mathrm{f}}=k_{\mathrm{s}}$, and obtain for the temperature field outside of the swimmer
\begin{align}
\label{eq:tsolout} t(\vec{r}) &= \frac{1}{2} \left( \frac{a}{r} \right) + \sum_{i=0}^{\infty} \bar{t}_{i} \left( \frac{a}{r} \right)^{i+1} P_{i} \left( \cos \theta \right)~, \\
\label{eq:tsolcoefo} \bar{t}_{i = 2k} &= \frac{1}{\pi}\frac{(-1)^{k}}{2k+1}~, \\
\label{eq:tsolcoefe} \bar{t}_{i = 2k + 1} &= -\frac{1}{\pi}\frac{(-1)^{k}}{2k+1}~,
\end{align}
where, $P_{i}$ is the $i$-th Legendre polynomial.

For a constant heat flux $Q$ into the cap, the temperature field reads
\begin{align}
\label{eq:tsolflux} T(\vec{r}) &= T^{\infty} + \frac{a Q}{2 k_{\mathrm{f}}} \left[ \left( \frac{a}{r} \right) + \sum_{i=0}^{\infty} \hat{t}_{i} \left( \frac{a}{r} \right)^{i+1} P_{i} \left( \cos \theta \right) \right]~, \\
\label{eq:tsolcoef_q_ie} \hat{t}_{i = 2k} &= 0~, \\
\label{eq:tsolcoef_q_io} \hat{t}_{i = 2k+1} &= - \frac{4k+3}{(2k + 2) + (2k + 1)K} \frac{ (-1)^{k}(2k)! }{ 2^{2k+1} k! (k+1)! }~,
\end{align}
where $K = k_{\mathrm{s}} / k_{\mathrm{f}}$ is the conductivity contrast. In this case, the maximum temperature difference appearing in our $\tau$ expansion can be written as
\begin{align}
\label{eq:deltaT_q} \Delta T &= \frac{a q}{2 k_{\mathrm{f}}} \left[ 1 - \sum_{i=0}^{\infty} \hat{t}_{i} \right]~,
\end{align}
leading to a reduced temperature field in a more convenient form for our purposes,
\begin{align}
\label{eq:tsoloutflux} t(\vec{r}) &= \left( \frac{a}{r} \right) \left[ 1 - \sum_{j=0}^{\infty} \hat{t}_{j} \right]^{-1} + \sum_{i=0}^{\infty} \bar{t}_{i} \left( \frac{a}{r} \right)^{i+1} P_{i} \left( \cos \theta \right)~, \\
\label{eq:tsoloutfluxcoef} \bar{t}_{i} &= \hat{t}_{i} \left[ 1 - \sum_{j=0}^{\infty} \hat{t}_{j} \right]^{-1}~.
\end{align}

\section{\label{app:linearized}The Equilibrium Solutions}

The linearized equations for the equilibrium in terms of our reduced quantities are as follows. The heat equation reduces to a constant temperature $T^{\infty}$ throughout the system. The Stokes equations reduce to zero fluid velocity, with the following pressure condition
\begin{align}
\label{eq:preseq} \vec{\nabla} p^{\mathrm{eq}}(\vec{r}) &= -k_{\mathrm{B}} T^{\infty} n^{\infty} \left( x_{+}^{\mathrm{eq}}(\vec{r}) - x_{-}^{\mathrm{eq}}(\vec{r}) \right) \vec{\nabla} \phi^{\mathrm{eq}}(\vec{r})~.
\end{align}
That is, the hydrostatic pressure exactly cancels the ionic pressure terms induced by electrostatic screening of any charge or potential on the colloid. The linearized equilibrium Poisson equation reads
\begin{align}
\label{eq:poislineq} \nabla^{2} \phi^{\mathrm{eq}}(\vec{r}) &= - \frac{1}{2} (\kappa^{\infty})^{2} \left( x_{+}^{\mathrm{eq}}(\vec{r}) - x_{-}^{\mathrm{eq}}(\vec{r}) \right)~.
\end{align}
Lastly, the ionic fluxes become
\begin{align}
\label{eq:fluxlineq} \vec{j}_{\pm}^{\mathrm{eq}}(\vec{r}) &= - D_{\pm}^{\infty} n^{\infty} \left[ \vec{\nabla} x_{\pm}^{\mathrm{eq}}(\vec{r}) \pm \vec{\nabla} \phi^{\mathrm{eq}}(\vec{r}) \right]~,
\end{align}
with the closure $\vec{j}_{\pm}^{\mathrm{eq}}(\vec{r}) = \vec{0}$. The latter follows from the fact that in equilibrium the fluxes vanish. Using the closure, we find that $x_{\pm}^{\mathrm{eq}}(\vec{r}) = \mp \phi^{\mathrm{eq}}(\vec{r})$ and $\nabla^{2} \phi^{\mathrm{eq}}(\vec{r}) = (\kappa^{\infty})^{2} \phi^{\mathrm{eq}}(\vec{r})$. The hydrostatic pressure condition reduces to $\vec{\nabla} p^{\mathrm{eq}}(\vec{r}) = 2 k_{\mathrm{B}} T^{\infty} n^{\infty} \phi^{\mathrm{eq}}(\vec{r}) \vec{\nabla} \phi^{\mathrm{eq}}(\vec{r})$.

\section{\label{app:teubner}The Expansion of Teubner's Integration}

We rewrite the expression for the speed given by Teubner~\cite{teubner1982motion}, see~\cref{eq:Tspeed}, in terms of the body force inside and outside the screening layer --- using that the latter is vanishing and that the system is axisymmetric --- to arrive at
\begin{align}
\label{eq:Usplit} \bar{U} &= \frac{1}{3 \eta^{\infty} a} \int_{a}^{\infty} r^{2} \int_{0}^{\pi} \sin \theta \underline{K}(\vec{r}) \cdot \vec{f}^{\mathrm{neq}}(\vec{r}) \mathrm{d} \theta \mathrm{d} r = \frac{1}{3 \eta^{\infty} a} \int_{a}^{a^{+}} r^{2} \int_{0}^{\pi} \sin \theta \underline{K}(\vec{r}) \cdot \vec{f}^{\mathrm{neq}}_{\mathrm{in}}(\vec{r}) \mathrm{d} \theta \mathrm{d} r~,
\end{align}
where $a^{+}$ marks the edge of the screening layer. We have that $\underline{K}(\vec{r}_{s}) = \vec{0}$ and we must therefore perform a perturbative analysis in terms of $\lambda^{\infty}/a \ll 1$ to obtain the relevant weighting factors over the length of the screening layer. We introduce $\vec{r} = \vec{r}_{s} + \lambda^{\infty} q \hat{q}$, such that 
\begin{align}
\label{eq:Uinnew} \bar{U} &= \frac{\lambda^{\infty}}{3 \eta^{\infty} a} \int_{0}^{\infty} \int_{0}^{\pi} \left(a + \lambda^{\infty} q \right)^{2} \sin \theta \underline{K}(\vec{r}_{s} + \lambda^{\infty} q \hat{q}) \cdot \vec{f}^{\mathrm{neq}}_{\mathrm{in}}(\vec{r}_{s} + \lambda^{\infty} q \hat{q}) \mathrm{d} \theta \mathrm{d} q~,
\end{align}
where the $\lambda^{\infty}$ term comes from the Jacobian of the coordinate transformation and we have taken the limit to infinity for the $q$ integration, as $a^{+} = \lim_{q \uparrow \infty} a + \lambda^{\infty} q$. The expressions for the terms that do not pertain to the force reduce to
\begin{align}
\label{eq:kernew} \frac{\lambda^{\infty}}{3 \eta^{\infty} a} \left(a + \lambda^{\infty} q \right)^{2} \sin \theta \underline{K}(\vec{r}_{s} + \lambda^{\infty} q \hat{q}) &= - \frac{\left( \lambda^{\infty} \right)^{3} q^{2}}{2 \eta^{\infty} a} \cos \theta \sin \theta \hat{q} + \frac{\left( \lambda^{\infty} \right)^{2} q}{2 \eta} \sin^{2} \theta \hat{\theta}~, & 
\end{align}
to leading order in $\lambda^{\infty}$. We verified that the next order terms do not contribute to the speed at leading order. For the force, we obtain
\begin{align}
\label{eq:forcerewrite} \vec{f}^{\mathrm{neq}}_{\mathrm{in}}(\vec{r}_{s} + \lambda^{\infty} q \hat{q}) &= A(q) \hat{q} t(a,\theta) + B(q) a^{-1} \hat{\theta} \partial_{\theta} t(a,\theta)~,
\end{align}
where the term $t(a,\theta) \equiv t(\vec{r}_{s})$ is the temperature at the surface. Here, $A(q)$ accounts for all the prefactors in~\cref{eq:fstinpersubspot,eq:fstinpersubscha}; $B(q)$ accounts for all the relevant prefactors in~\cref{eq:fstinparpot,eq:fstinparcha}; and $a^{-1} \hat{\theta} \partial_{\theta} t(a,\theta) \equiv \vec{\nabla}_{\parallel} t(\vec{r}_{s})$. Note that here we have used our assumption that $\phi(\vec{r}_{s})$ is homogeneous over the surface to avoid $q$ dependence in the factors $A$ and $B$.

Now taking everything together, we may rewrite the expression for the speed contribution due to the region inside of the thin screening layer as
\begin{align}
\label{eq:Uinrew} \bar{U} &\approx \frac{\left( \lambda^{\infty} \right)^{2}}{2 \eta^{\infty} a} \int_{0}^{\infty} \int_{0}^{\pi} \left[ \lambda^{\infty} q^{2} \cos \theta \sin \theta A(q) t(a,\theta) - q \sin^{2} \theta B(q) \partial_{\theta} t(a,\theta) \right] \mathrm{d} \theta \mathrm{d} q~.
\end{align}
Spitting the integrand into the $A(q)$ ($\perp$) and $B(q)$ ($\parallel$) terms, we evaluate these contributions separately. Starting with the perpendicular component, we find
\begin{align}
\nonumber \bar{U}_{\perp} &= \frac{\left( \lambda^{\infty} \right)^{3}}{2 \eta^{\infty} a} \int_{0}^{\infty} \int_{0}^{\pi} q^{2} \cos \theta \sin \theta A(q) t(a,\theta) \mathrm{d} \theta \mathrm{d} q \\
\label{eq:Uinrewperp} &= \frac{\left( \lambda^{\infty} \right)^{3}}{2 \eta^{\infty} a} \int_{0}^{\infty} q^{2} A(q) \mathrm{d} q \int_{0}^{\pi} t(a,\theta) \cos \theta \sin \theta \mathrm{d} \theta = \frac{\left( \lambda^{\infty} \right)^{3}}{3 \eta^{\infty} a} \bar{t}_{1} \int_{0}^{\infty} q^{2} A(q) \mathrm{d} q~,
\end{align}
where only the first-order Legendre Polynomial contributes. Note that if we had not assumed homogeneous electrostatic surface properties, the splitting of the integration could not have been done in the same way and all Legendre-Fourier modes would have contributed. Evaluating the integral over $q$ gives us for a conducting surface
\begin{align}
\label{eq:Uinperpfinpot} \bar{U}_{\perp} &= - \frac{k_{\mathrm{B}} T^{\infty} n^{\infty}}{6 \eta^{\infty} a} \left( \lambda^{\infty} \right)^{2} \left[ 2 \beta \phi_{0} + \epsilon^{*} \phi_{0}^{2} \right] \bar{t}_{1}~.
\end{align}
The result for an insulating surface is 
\begin{align}
\label{eq:Uinperpfincha} \bar{U}_{\perp} &= - \frac{k_{\mathrm{B}} T^{\infty} n^{\infty}}{6 \eta^{\infty} a} \left( \lambda^{\infty} \right)^{2} \left[ 1 + 2 \epsilon^{*} \right] \phi_{s}^{2} \bar{t}_{1}~.
\end{align}

Similarly, we obtain for the parallel component 
\begin{align}
\nonumber \bar{U}_{\parallel} &= \frac{\left( \lambda^{\infty} \right)^{2}}{2 \eta^{\infty} a} \int_{0}^{\infty} \int_{0}^{\pi} q \sin^{2} \theta B(q) \partial_{\theta} t(a,\theta) \mathrm{d} \theta \mathrm{d} q \\
\label{eq:Uinrewpara} &= \frac{\left( \lambda^{\infty} \right)^{2}}{2 \eta^{\infty} a} \int_{0}^{\infty} q B(q) \mathrm{d} q \int_{0}^{\pi} \sin^{2} \theta \partial_{\theta} t(a,\theta) \mathrm{d} \theta = - \frac{2 \left( \lambda^{\infty} \right)^{2}}{3 \eta^{\infty} a} \bar{t}_{1} \int_{0}^{\infty} q B(q) \mathrm{d} q~.
\end{align} 
Evaluating the integral over $q$ leads to the desired expression for a conducting surface
\begin{align}
\label{eq:Uinparafinpot} \bar{U}_{\parallel} &= - \frac{k_{\mathrm{B}} T^{\infty} n^{\infty}}{6 \eta^{\infty} a} \left( \lambda^{\infty} \right)^{2} \left[ 6 \beta \phi_{0} - \phi_{0}^{2} \right] \bar{t}_{1}~,
\end{align} 
and for an insulating surface
\begin{align}
\label{eq:Uinparafinchar} \bar{U}_{\parallel} &= - \frac{k_{\mathrm{B}} T^{\infty} n^{\infty}}{6 \eta^{\infty} a} \left( \lambda^{\infty} \right)^{2} \left[ 8 \beta \phi_{s} - \left( 2 + \epsilon^{*} \right) \phi_{s}^{2} \right] \bar{t}_{1}~.
\end{align} 

\end{document}